\documentclass[%
superscriptaddress,
preprint,
 amsmath,amssymb,
 aps,
 prd,
]{revtex4-1}
\usepackage[colorlinks=true,linkcolor=blue]{hyperref}
\usepackage{graphicx}
\usepackage{dcolumn}
\usepackage{bm}
\usepackage{subfigure}

\begin{document}

\title{$U(1)$ Gauge Field Localization on Bloch Brane with Chumbes-Holf da Silva-Hott Mechanism}

\author{Zhen-Hua Zhao}
\email{zhaozhh09@lzu.edu.cn}
\affiliation{$^a$Department of Applied Physics,
    Shandong University of Science and Technology,
    Qingdao 266590, People's Republic of China}
\author{Yu-Xiao Liu}
\email{liuyx@lzu.edu.cn}

\affiliation{Institute of Theoretical Physics,
        Lanzhou University, Lanzhou 730000,
        People's Republic of China}
\author{Yuan Zhong}
\email{yzhong@ifae.es}
\affiliation{Institute of Theoretical Physics,
        Lanzhou University, Lanzhou 730000,
        People's Republic of China}
\affiliation{IFAE, Universitat Aut\`onoma de Barcelona, 08193 Bellaterra, Barcelona, Spain}

\begin{abstract}{We follow the Chumbes-Holf da Silva-Hott (CHH) mechanism to study the (qusi-)localization of $U(1)$ gauge field on the Bloch brane.  {The localization and resonances of $U(1)$ gauge field are discussed for four kinds of Bloch brane solutions: the original and generalized Bloch brane solutions, as well as the degenerate Bloch brane solutions I and II. With the CHH mechanism, we find that the mass spectrum of the gauge field Kaluza-Klein modes is continuous and there is no the tachyonic mode. The zero mode is localized on all the branes and there are resonant KK modes on the degenerate Bloch branes.}
}
\end{abstract}


 \maketitle

\section{Introduction}
In brane-world theory  graviton can be localized on the Randall-Sundrum (RS) thin  brane \cite{Randall199983,Randall199983a} and on the thick
brane \cite{Gremm2000478}, naturally. In addition to graviton localization, the localization of Standard Model particles is also an important issue for any brane-world model.

The localization of fermion on brane can be realized by introducing the usual Yukawa coupling between the background scalar field and the fermion field \cite{Rubakov1983,Bajc2000,Randjbar-Daemi2000,Ringeval200265,Koley200522,Melfo2006,LiuZhangZhangDuan2008} when the background scalar field has a kink-like configuration interpolating between the two different vacua
at the two sides of brane. However, when the scalar field is an even function of the extra dimension, one need to introduce the new localization mechanism presented in Ref.~\cite{LiuXuChenWei2013}. Real scalar field can be localized on brane as long as graviton is localizable~\cite{Bajc2000}.

While for the $U(1)$ gauge field, its localization is more complex than the fermion and scalar fields.
In the RS thin  brane scenario, $U(1)$ gauge field
  with the following standard five-dimensional action
\begin{equation}
S\sim\int d^5x\sqrt{-g}F_{MN}F^{MN} \label{action0}
\end{equation}
can not be localized \cite{Pomarol2000}. Here, $F_{MN}=\partial_M A_N-\partial_N A_M$ is the field strength of the $U(1)$ gauge field. In order to localize $U(1)$ gauge field on the RS brane, many ideas were proposed  \cite{Oda2001,Oda2000,Dvali2001,Dimopoulos2001,Guerrero2009,Ghoroku2002,CarenaPontonTaitWagner2003,DavoudiaslHewettRizzo2003a,Giovannini2002}.

In the thick brane scenario, $U(1)$ gauge field with the action (\ref{action0}) can be localized on some thick branes. For example, it can be localized on the thick de Sitter ($dS$) brane \cite{Liu20090902,GuoHerrera-AguilarLiuMalagon-MorejonMora-Luna2013,Herrera-AguilarRojasSantos-Rodriguez2014}, the Weyl thick brane  \cite{Liu200808}, and the brane with finite extra-dimension \cite{LiuFuGuoLi2012}. Especially, in Ref.~\cite{GuoHerrera-AguilarLiuMalagon-MorejonMora-Luna2013}, there the potentials in the corresponding Schrodinger equations for the  Kaluza-Klein (KK) modes of the vector field are modified Poschl-Teller potentials, which lead to the localization of the vector zero mode on the brane as well as to mass gaps in the mass spectra. But it cannot be localized on thick brane models that are asymptotically RS.

In order to localize gauge field on thick brane, Kehagias and Tamvakis (KT)
proposed a general mechanism, in which a coupling between the gauge field and an extra dilaton field is introduced  \cite{Kehagias2001504}. Kehagias and
Tamvakis mechanism has been applied in many different
brane-world scenarios to localize the vector
\cite{CruzTahimAlmeida2010,AlencarLandimTahimMunizCosta2010,CruzLimaAlmeida2013,FuLiuGuo2011} and Kalb-Ramond fields \cite{CruzTahimAlmeida2009,ChristiansenCunhaTahim2010,CruzMalufAlmeida2013,FuLiuGuo2011}.
Recently, Chumbes, Holf da Silva and Hott (CHH) propose a new mechanism to localize gauge and tensor fields on thick brane \cite{ChumbesHoffHott2012}. In their method, gauge and tensor fields are directly coupled to a function of the background scalar field.

On the other hand, the thick brane is usually generated by a background scalar field. In Ref. \cite{Bazeia200405}, Bazeia and Gomes introduced the Bloch brane generated by two real scalar fields. This brane model was further generalized in Ref. \cite{SouzaDutra200878}, and investigated in Refs.~\cite{Gomes2006, CorreaDutraHott2010,  CruzLimaAlmeida2013, CruzMalufAlmeida2013,  XieYangZhao2013,  LiuXuChenWei2013}. It is known that $U(1)$ gauge field with the action (\ref{action0}) can not be localized on the Bloch brane \cite{CruzLimaAlmeida2013}. In Ref. \cite{CruzLimaAlmeida2013} the localization of guage field on the Bloch brane was discussed with the KT mechanism. In order to localize the zero mode on the Bloch brane, an extra dilaton scalar field is introduced in Ref. \cite{CruzLimaAlmeida2013}.

In this paper, we will investigate the localization of $U(1)$ gauge field with the CHH mechanism.
In this mechanism the third dilaton scalar field, which appears in Ref. \cite{CruzLimaAlmeida2013}, is not needed.  {The localization and quasilocalization of gauge field for four kinds of Bloch brane solutions are discussed and the localized zero mode and resonant KK modes are found.}

This paper is constructed as follows. In section \ref{BB}, the Bloch brane scenario and its four kinds of solutions are reviewed briefly. The localization and quasilocalization of $U(1)$ gauge field are discussed in section \ref{LocGF}. Finally, we give our conclusions in section \ref{Cons}.

\section{Review of Bloch brane} \label{BB}

The action for the Bloch brane model reads \cite{Bazeia200405}
\begin{eqnarray}
S=&&\int d^4xdy\sqrt{-g}\bigg[\frac{1}{4}R-\frac{1}{2}\partial_M\phi\partial^M\phi
   -\frac{1}{2}\partial_M\chi\partial^M\chi\nonumber\\
 &&   -V(\phi,\chi)
   \bigg]\label{action1},
\end{eqnarray}
where $g=\textrm{det}(g_{MN})$, $R$ is the scalar curvature of the five-dimensional space-time, $M,N=0,1,2,3,4$, and $\phi$, $\chi$ are two real scalar fields depending only on the extra-dimensional coordinate $y$ for the static flat brane model.

The line element for the five-dimensional space-time is assumed as
\begin{eqnarray}
ds^2={g}_{MN}dx^Mdx^N=e^{2\alpha(y)}\eta_{\mu\nu}dx^{\mu}dx^{\nu}+dy^2,\label{metric}
\end{eqnarray}
where $e^{2 \alpha(y)}$ is the warp factor, $\alpha(y)$ is only the function of extra-dimensional coordinate $y$, and $\eta_{\mu\nu}=\textrm{diag}{(-1,1,1,1)}$.
From the above action (\ref{action1}), one can get the equations of motion of $\phi$, $\chi$, and
the Einstein equations \cite{Bazeia200405}:
\begin{eqnarray}
\phi''&=&-4\alpha'\phi'+\frac{\partial V(\chi,\phi)}{\partial \phi},\label{eq11}\\
\chi''&=&-4\alpha'\chi'+\frac{\partial V(\chi,\phi)}{\partial \chi},\label{eq12}\\
\alpha''&=&-\frac{2}{3}(\phi'^2+\chi'^2),\label{eq13}\\
\alpha'^2&=&\frac{1}{6}(\phi'^2+\chi'^2)-\frac{1}{3}V(\phi,\chi),\label{eq14}
\end{eqnarray}
where the prime stands for the derivative with respect to $y$.
By introducing a superpotential $W(\phi,\chi)$, the above equations can be reduced to the following first-order form:
\begin{eqnarray}
\phi'&=&\frac{\partial W(\chi,\phi)}{\partial \phi},\label{phi}\\
\chi'&=&\frac{\partial W(\chi,\phi)}{\partial \chi},\label{chi}\\
\alpha'&=&-\frac{2}{3}W(\chi,\phi),\label{alpha}
\end{eqnarray}
and the scalar potential is determined in terms of the superpotential by
\begin{eqnarray}
V = &&\frac{1}{2}\left[ \left(\frac{\partial W(\chi,\phi)}{\partial\phi}\right)^2 +\left(\frac{\partial W(\chi,\phi)}{\partial\chi}\right)^2 \right] \nonumber\\
&&-\frac{4}{3}W^2(\chi,\phi).
\end{eqnarray}
Then for the superpotential
\begin{equation}
W(\phi,\chi)=\phi\left[\left(1-\frac{1}{3}\phi^2\right)-b\chi^2\right], \label{SuperPotential1}
\end{equation}
where $b$ is a real parameter,
the solution of Eqs.~(\ref{phi})-(\ref{alpha}) is given by \cite{Bazeia200405}
\begin{subequations}\label{BlochSolution1}
\begin{eqnarray}
\phi(y)&=&\tanh(2by),\label{s11}\\
\chi(y)&=&\sqrt{\frac{1}{b}-2}\;\textrm{sech}(2by),\label{s12}\\
\alpha(y)&=&\frac{1}{9b}[(1-3b)\tanh^2(2by)\nonumber\\
& &-2\ln\cosh(2by)],\label{s13}
\end{eqnarray}
\end{subequations}
where the parameter $b$ satisfies the constrain $0<b<1/2$.  The above two-field solution represents a Bloch wall. When $b\rightarrow 1/2$, one will get the Ising wall of the one-field solution \cite{Bazeia200405}.

In addition to the above original Bloch brane solution, the generalized Bloch brane solution was found in Ref.~\cite{SouzaDutra200878} by using the following generalized superpotential
\begin{equation}
W(\phi,\chi)=\phi\left[a\left(v^2-\frac{1}{3}\phi^2\right)-b\chi^2\right],  \label{SuperPotential2}
\end{equation}
 {where $a$, $b$, and $v$ are positive constants.}
It reads \cite{SouzaDutra200878}
\begin{subequations}\label{BlochSolution2}
\begin{eqnarray}
\phi(y)&=&v\tanh(2bvy),\label{s21}\\
\chi(y)&=&v\sqrt{\frac{a-2b}{b}}\;\textrm{sech}(2bvy),\label{s22}\\
\alpha(y)&=&\frac{v^2}{9b}[(a-3b)\tanh^2(2bvy)\nonumber\\
& &-2a\ln\cosh(2bvy)],\label{s23}
\end{eqnarray}
\end{subequations}
where $a>2b>0$.

Other solutions of Bloch brane were also found in Ref.~\cite{SouzaDutra200878} for the same superpotential (\ref{SuperPotential2}) with $a=b$ and $a=4b$, namely, the degenerated Bloch brane solutions.
They are
\begin{subequations}\label{BlochSolution3}
\begin{eqnarray}
\phi(y)&=& \frac{\sqrt{c_0^2-4}~ v \sinh (2 b v y)}{\sqrt{c_0^2-4} \cosh (2 b v y)-c_0},\label{s311}\\
\chi(y)&=& \frac{2 v}{\sqrt{c_0^2-4} \cosh (2 b v y)-c_0},\label{s312}\\
\alpha(y)&=&\frac{1}{2} \Bigg[\frac{4 v^2 \left(-\sqrt{c_0^2-4} ~c_0 \cosh (2 b v
   y)+c_0^2-4\right)}{9 \left(\sqrt{c_0^2-4} \cosh (2 b v y)-c_0\right)^2}\nonumber\\
   & &-\frac{4
   \left(c_0^2-\sqrt{c_0^2-4} c_0-4\right) v^2}{9
   \left(\sqrt{c_0^2-4}-c_0\right)^2}\Bigg]\nonumber\\
   &&+\frac{1}{2} \log
   \left(\frac{\sqrt{c_0^2-4}-c_0}{\sqrt{c_0^2-4} \cosh (2 b v
   y)-c_0}\right)^{\frac{4 v^2}{9}}, \label{s313}
\end{eqnarray}
\end{subequations}
for $c_0<-2$ and $a=b$,
and
\begin{subequations}\label{BlochSolution4}
\begin{eqnarray}
\phi(y)&=& \frac{\sqrt{1-16 c_0} v \sinh (4 b v y)}{\sqrt{1-16 c_0} \cosh (4 b v y)+1},\label{s321}\\
\chi(y)&=& \frac{2 v}{\sqrt{\sqrt{1-16 c_0} \cosh (4 b v y)+1}},\label{s322}\\
\alpha(y)&=&\frac{1}{2} \Bigg[\frac{4 \left(8 c_0+\sqrt{1-16 c_0}+1\right) v^2}{9 \left(\sqrt{1-16
   c_0}+1\right)^2}\nonumber\\
   &&-\frac{4 v^2 \left(\sqrt{1-16 c_0} \cosh (4 b v y)+8
   c_0+1\right)}{9 \left(\sqrt{1-16 c_0} \cosh (4 b v y)+1\right)^2}\Bigg]\nonumber\\
   &&+\frac{1}{2}
   \log\left(\frac{\sqrt{1-16 c_0}+1}{\sqrt{1-16 c_0} \cosh (4 b v
   y)+1}\right)^{\frac{8 v^2}{9}},\label{s323}
\end{eqnarray}
\end{subequations}
for $c_0<1/16$ and $a=4b$.

In this paper we will call solutions (\ref{BlochSolution1}), (\ref{BlochSolution2}), (\ref{BlochSolution3}), and (\ref{BlochSolution4}) as the original,
generalized, degenerate I, and degenerate II Bloch brane solutions, respectively.

From the above solutions one can find that the Bloch brane has  a rich inner structure.  The details of the above solutions can be fund in Refs.  \cite{Bazeia200405,SouzaDutra200878}.

\section{Localization and quasilocalization of gauge field }\label{LocGF}

As was analyzed in Ref. \cite{CruzLimaAlmeida2013}, for a gauge field with the following standard five-dimensional action
\begin{equation}
S\sim\int d^5x\sqrt{-g}F_{MN}F^{MN},
\end{equation}
the corresponding zero mode cannot be localized on the Bloch brane.
In order to localize gauge field on the Bloch brane, the authors of Ref. \cite{CruzLimaAlmeida2013} extended the Bloch brane scenario to the so called dilatonic Bloch brane model, which is described by the following action 
\begin{eqnarray}
S=&&\int d^5x\sqrt{-g}\bigg[\frac{1}{4}R
   -\frac{1}{2}(\partial\phi)^2
   -\frac{1}{2}(\partial\chi)^2 -\frac{1}{2}(\partial\pi)^2\nonumber\\
   &&-V(\phi,\chi,\pi)
   \bigg], \label{action2}
\end{eqnarray}
where the scalar fields $\phi$ and $\chi$ generate the brane, and the dilaton scalar field $\pi$ is used to localize gauge field on the brane.
The action of gauge field is assumed to be
\begin{equation}
S\sim\int \sqrt{-g} d^5x \,\text{e}^{-2\lambda\pi\sqrt{2/3}}F_{MN}F^{MN}\label{action3},
\end{equation}
where the coupling between the dilation field $\pi$ and the gauge field is introduced. With the action (\ref{action3}), the zero mass mode of gauge field was found to be localized on the brane and some massive resonant modes were also found \cite{CruzLimaAlmeida2013}.

The method used in Ref. \cite{CruzLimaAlmeida2013} was first proposed by Kehagias and Tamvakis (KT) in Ref. \cite{Kehagias2001504}.
In this paper we will follow another mechanism proposed by Chumbes, Hoff, and Hott (CHH) in Ref. \cite{ChumbesHoffHott2012} and study the localization and quasilocalization of gauge field. Compared with the KT mechanism, the dilation scalar field is not needed in the CHH mechanism. We will show that $U(1)$ gauge field can be localized on the standard Bloch brane by introducing a coupling
between the gauge field and the background scalar field $\chi$. The action of the five-dimensional $U(1)$ gauge field reads
\begin{equation}
S=-\frac{1}{4}\int d^5x\sqrt{-g}\chi(y) F_{MN}F^{MN}\label{action4},
\end{equation}
By means of the decomposition of $A_{\mu}=\sum_n a_{\mu}(x)\rho_n(y)$ and the gauge $\partial_{\mu} A^{\mu}=0$ and $A_4=0$, the above action (\ref{action4}) can be reduced to
\begin{equation}
S=-\frac{1}{4}\int dy \chi(y)\rho_n(y)^2\int d^4x(f_{\mu\nu}f^{\mu\nu}-2 m_n^2 a_{\mu}a^{\mu})\label{action5},
\end{equation}
where $f_{\mu\nu}=\partial_{\mu}a_{\nu}-\partial_{\nu}a_{\mu}$ is the four-dimensional gauge field strength tensor, and $\rho_n(y)$ should satisfy the equation
\begin{equation}
\rho_n''+\left(\frac{\chi'}{\chi}+2\alpha'\right)\rho_n'=-m_n^2\rho_n e^{-2\alpha} \label{eq}.
\end{equation}
 The localization of gauge field requires
\begin{equation}
I\equiv\int_{-\infty}^{+\infty} dy \chi(y)\rho_n^2(y)<\infty \label{int}.
\end{equation}

\subsection{Zero mode}

First we discuss the localization of the zero mode of gauge field. Let $m_0=0$, Eq.~(\ref{eq}) reads
\begin{equation}
\rho_0''+\left(\frac{\chi'}{\chi}+2\alpha'\right)\rho_0'=0. \label{eq0}
\end{equation}
By introducing the filed transformation  \cite{ChumbesHoffHott2012}
\begin{equation}
\rho_0=e^{-\gamma(y)}\hat{\rho}_0(y)
\end{equation}
with $\gamma(y)$ satisfies $2\gamma'=2\alpha+\chi'/\chi$,
Eq.~(\ref{eq0}) can be reduced to
\begin{equation}
-\hat{\rho}_0''+\left(\gamma''+\gamma'^2\right)\hat{\rho}_0=0, \label{eq1}
\end{equation}
or
\begin{equation}
\left(\frac{d}{dy}+\gamma'\right)\left(-\frac{d}{dy}+\gamma'\right)\hat{\rho}_0=0. \label{eq2}
\end{equation}
The solution of the above equation is $\hat{\rho}_0=C_1 e^{\gamma(y)}$, where $C_1$ is a constant. So the zero mode solution is ${\rho}_0=C_1$.
 {
The localization of the zero mode needs
\begin{eqnarray}
I&=&\int_{-\infty}^{+\infty} dy \chi(y)\rho_0^2(y)\nonumber\\
&=&C_{1}^{2}\int_{-\infty}^{+\infty}  \chi(y) dy<\infty \label{int2}.
\end{eqnarray}
Because the function $\chi(y)$ is continuous, the convergence of the above integration is decided by the asymptotic behavior of $\chi(y)$ at the infinity of extra dimension. }

 {
In addition to the above four kinds of analytic solutions, there may exist some other solutions of the Bloch brane. And to check the localization condition (\ref{int2}) for all of these brane solutions one by one is not efficient. Therefore, in the following we will try to find out the general asymptotic solution of $\chi(y)$ at the infinity and give a general conclusion.}

 {
In the Bloch brane scenario, the configuration of the scalar $\phi(y)$ is a kink, and its asymptotic solution is
\begin{eqnarray}
\phi(y\rightarrow\pm \infty)&\rightarrow& \pm v,  \label{asympphi}
\end{eqnarray}
where $v$ is the vacuum expectation value of $\phi$.
Substituting the general superpotential \eqref{SuperPotential2} into Eq. \eqref{chi} yields
\begin{eqnarray}
\chi'&=&-2b\phi\chi .\label{chi2}
\end{eqnarray}
When $y\rightarrow\pm\infty$, we have
\begin{eqnarray}
\chi'(y\rightarrow\pm\infty)&\rightarrow&\mp 2b v\chi(y\rightarrow\pm\infty),\label{chi3}
\end{eqnarray}
from which we can obtain the asymptotic solution of $\chi$:
\begin{eqnarray}
\chi(y\rightarrow\pm\infty)&\rightarrow&e^{-2b v|y|}.\label{chi4}
\end{eqnarray}
With the above asymptotic solution, we know that the integration in Eq. \eqref{int2} is convergent because the constants $b$ and $v$ are positive.
So the zero mode of vector field can be localized on the general Bloch brane.
}

 {Now we will calculate explicitly the  normalization constant $C_1$ for the four kinds of Bloch brane solutions.}

 {For the original and generalized Bloch brane solutions (\ref{BlochSolution1}) and (\ref{BlochSolution2}),
the  normalization constants are respectively given by
\begin{eqnarray}
C_1&=& \sqrt{\frac{2 b}{\pi}\sqrt{\frac{b}{1-2b}}}, ~~~(0<b<1/2)\\
C_1&=& \sqrt{\frac{2 b}{\pi}\sqrt{\frac{b}{a-2b}}},  ~~~(0<b<a/2)
\end{eqnarray}
which are finite.}

 {
For the degenerate Bloch solutions I (\ref{BlochSolution3}) and II (\ref{BlochSolution4}),
the results are respectively
\begin{eqnarray}
C_1 &=& \left[- \frac{2}{b} {\text{arctanh}
                 \left(\frac{1}{2} \big(\sqrt{c_0^2-4}+c_0\big)\right)}
        \right]^{-\frac{1}{2}},
(c_0<-2)\\
C_1 &=& \left[\frac{\sqrt{\sqrt{1-16 {c_0}}+1}}
             {4 K\left(1-\frac{2 \sqrt{1-16 {c_0}}}{\sqrt{1-16 {c_0}}+1}\right)} 
        \right]^{\frac{1}{2}}, ~~~(c_0<1/16)\label{DB2}
\end{eqnarray}
where the function $K(x)$ \footnote{The details of this function can be found in the website: http://mathworld.wolfram.com/CompleteEllipticIntegraloftheFirstKind.html}
gives the complete elliptic integral of the first kind. }

To sum up, with the CHH mechanism, the zero mode of $U(1)$ gauge field can be localized on the Bloch brane.

\subsection{Massive modes}
Next we investigate the (quasi)localization of the massive modes of gauge field.
In this part, it is more convenient to rewrite the metric \eqref{metric} in a  {conformally flat} form,
namely,
\begin{equation}
ds^2=e^{2A(z)}(\eta_{\mu\nu}dx^{\mu}dx^{\nu}+dz^2).\label{gcf}
\end{equation}
With the above metric (\ref{gcf}) and the gauge choice $A_4=0$, the action of the five-dimensional gauge field (\ref{action4})
is reduced to
\begin{equation}
S=-\frac{1}{4}\sum_{n}\int dz \tilde{\rho}^2_n(z)\int d^4x(f^{(n)}_{\mu\nu}f^{(n)\mu\nu}-2 m_n^2 a^{(n)}_{\mu}a^{{(n)}\mu})\label{action7},
\end{equation}
where $\tilde{\rho}_n=\rho_n \chi^{1/2}e^{A}$, and $\tilde{\rho}_n(z)$ satisfies the following Schr\"odinger-like equation
\begin{equation}
-\tilde{\rho}''_n+V(z)\tilde{\rho}_n= m_n^2\tilde{\rho}_n, \label{eq3}
\end{equation}
where the effective potential is given by
\begin{equation}
V(z)=\frac{1}{2}A''(z)+\frac{1}{4}
   A'^2(z)+\frac{A'(z) \chi'(z)}{2 \chi(z)}+\frac{\chi''(z)}{2 \chi(z)}-\frac{\chi'^2(z)}{4 \chi^2(z)},\label{V1}
\end{equation}
and the prime denotes the derivative with respect to $z$.
The above equation (\ref{eq3}) can be recast to
\begin{equation}
\mathcal{T}^\dagger\mathcal{T}\tilde{\rho}_n=m_n^2\tilde{\rho}_n, \label{eq4}
\end{equation}
where
\[
\mathcal{T}^\dagger=-\frac{d}{dz}+\Gamma,\;\;\mathcal{T}=\frac{d}{dz}+\Gamma,
\]
and
\[\Gamma=-\frac{1}{2}\left(\frac{\chi'}{\chi}+A'\right).\]
Equation \eqref{eq4} means that there is no tachyonic mode with $m^2<0$ in the spectrum of the KK modes \cite{Bazeia2004}.
Note that, in order to get the effective action (\ref{action7}) of the four-dimensional gauge fields from the five-dimensional one (\ref{action4}), we have introduced the orthonormalization condition between different massive  modes:
\begin{equation}
\int dz \tilde{\rho}_m(z) \tilde{\rho}_n(z) = 0. ~~~(m\neq n) 
\end{equation}
So the localization condition for $\tilde{\rho}_n(z)$ is
\begin{equation}
\int dz \tilde{\rho}^2_n(z)<\infty.
\end{equation}

The property of $\tilde{\rho}_n$ is determined by the effective potential $V(z)$ in (\ref{V1}). There are two methods to get the explicit expression of the effective potential. The first one is to
resolve the Einstein equations and the equations of motion of the background scalar fields
with the line element (\ref{gcf}). From our knowledge, with this method, there is no analytic
solution. The second one is to write the expression of $V(z(y))$ in the $y$ coordinate by the use of the coordinate transformation $dz=\text{e}^{-\alpha(y)}dy$, and the result is
\begin{eqnarray}
V(z(y))=&&\frac{1}{2} e^{2 \alpha(y)} \bigg(\alpha''(y)+\frac{2 \alpha'(y) \chi '(y)}{\chi (y)}+\frac{3}{2} \alpha'^2(y)\nonumber\\
&&+\frac{\chi''(y)}{\chi (y)}-\frac{\chi '^2(y)}{2 \chi^2 (y)}\bigg).\label{Vy}
\end{eqnarray}
Then, we can use the numerical relation between $y$ and $z$, $y=y(z)$, to obtain $V(z)$ from \eqref{Vy}.



\begin{figure*}
\begin{center}
\subfigure[$m^2=0.2$]{\label{figOBUzWaveFunction1}
\includegraphics[width=0.45\textwidth]{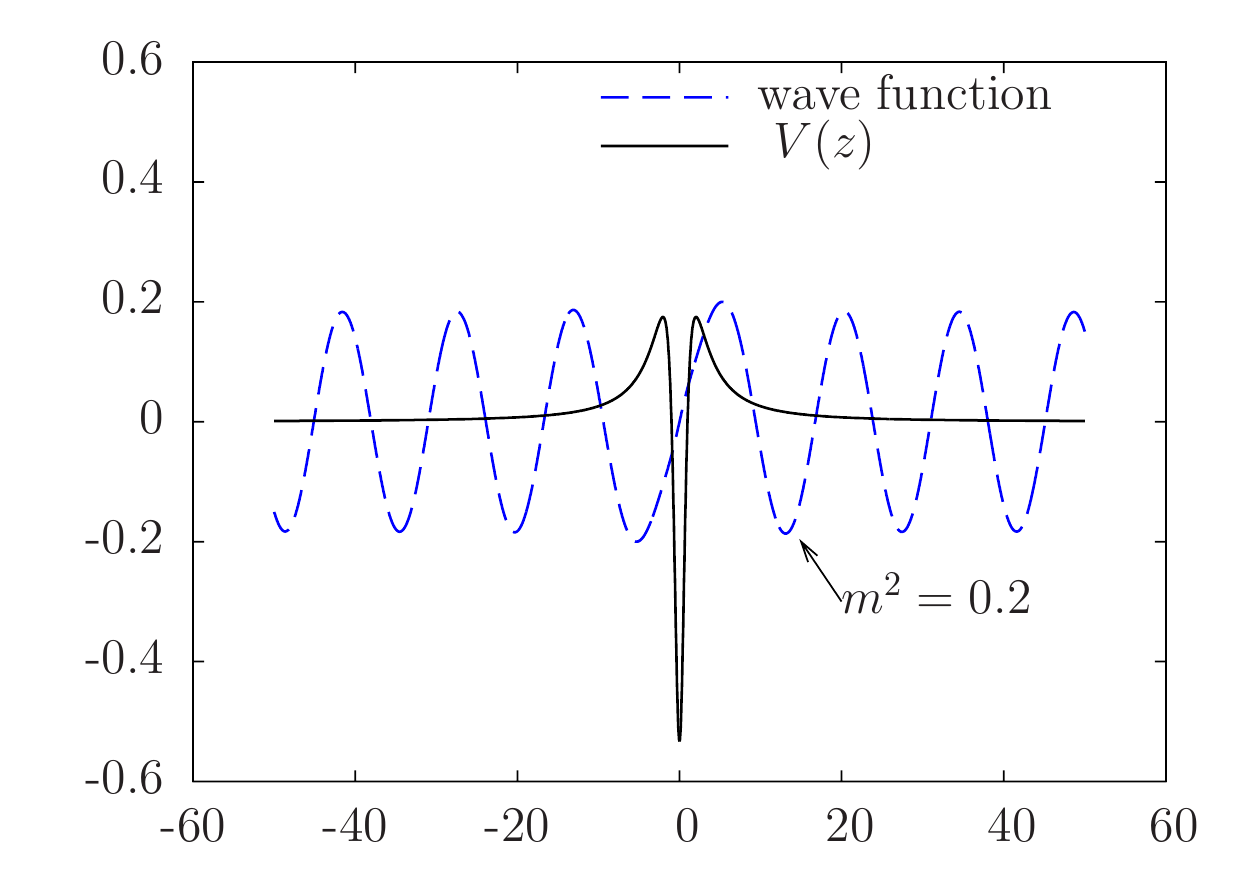}}
\subfigure[$m^2=1$]{\label{figOBUzWaveFunction2}
\includegraphics[width=0.45\textwidth]{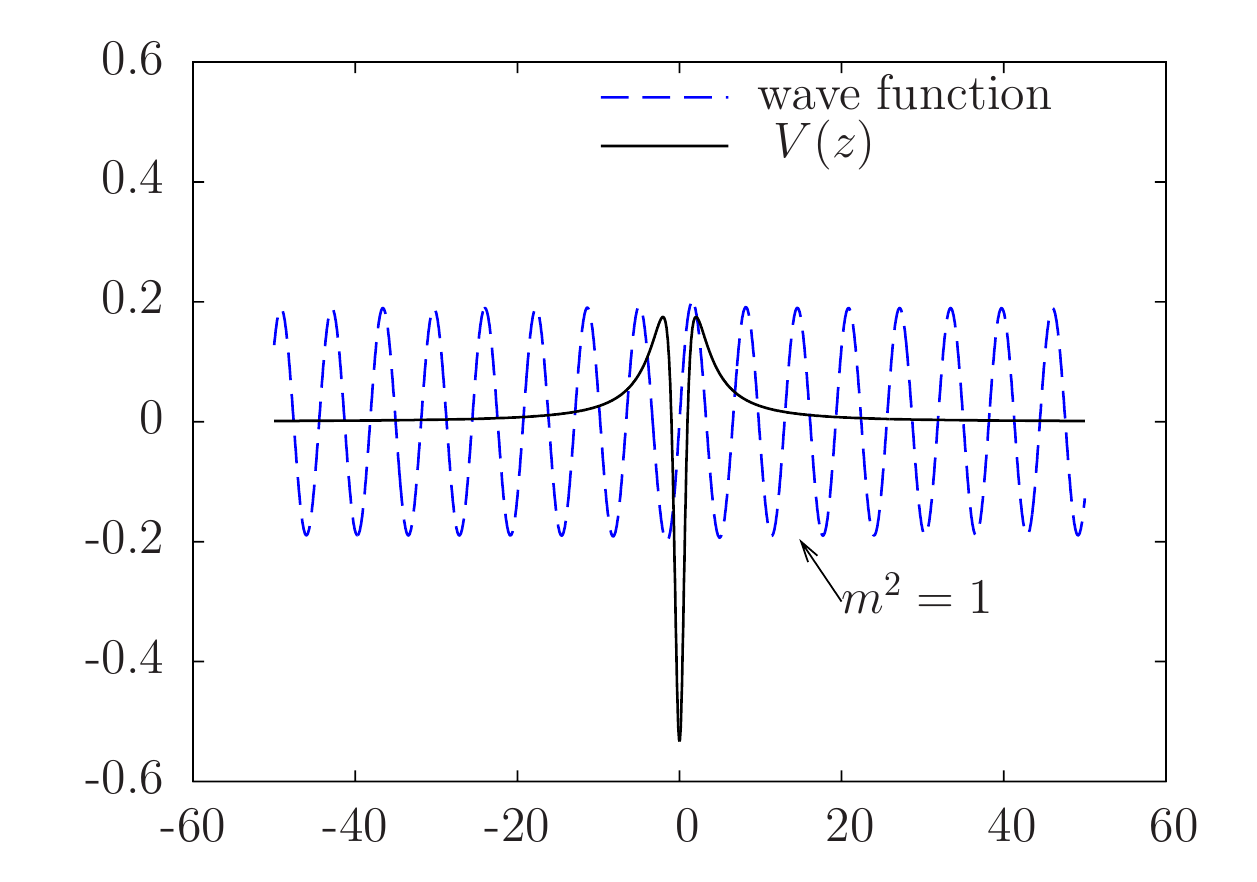}}
\caption{The shapes of the effective potential $V(z)$ and  {$\tilde{\rho}_n(z)$} for the original Bloch brane solution. The parameters are set to for $b=0.4$, $m^2=0.2$ (left), and $m^2=1$ (right). \label{VzWavefun}}
\end{center}
\end{figure*}

For all the Bloch brane solutions, the effective potentials $V(z)$ are of volcano-type and they trend to vanish at the infinity of extra dimension, i.e.,
\begin{equation}
V(z\rightarrow\pm\infty)\rightarrow 0.
\end{equation}
So the mass spectrum of the gauge field is continued and $m\geq 0$.  { For the massive KK mode, the solution of $\tilde{\rho}_n(z)$ oscillates when far away from the brane along the extra dimension. And when $m^2\gg V_{\rm max}$,
where $V_{\rm max}$ is the maximum of the $V(z)$, $\tilde{\rho}_n(z)$ approaches to the plane wave solution.  The shapes of $\tilde{\rho}_n(z)$} are shown in Fig.~\ref{VzWavefun} for a typical potential.

Since the effective potential trends to vanish at the boundary of the extra dimension, the massive KK modes cannot be normalized.

 {
In order to investigate the structure of the mass spectrum of these nonlocalized KK modes, we use the relative probability method introduced in Ref. \cite{Liu2009}. The relative probability function is defined as \cite{Liu2009}
\begin{equation}
P(m)=\frac{\int_{-z_b}^{z_b}\tilde{\rho}^2(z) dz}{\int_{-z_c}^{z_c}\tilde{\rho}^2(z) dz},\label{P}
\end{equation}
where $z_c>z_b$ and $2 z_b$ is of about the thickness of the brane. Here and after, we set $z_c=10z_b$. If $m^2\gg V_{\rm max}$ the solution of $\tilde{\rho}$ will be approximately a plane wave, so $P(m)\approx z_b/z_c=0.1$. In order to get the solution of Eq.~(\ref{eq3}), we introduce two boundary conditions:
\begin{equation}
\tilde{\rho}(0)=0,\quad \tilde{\rho}'(0)=1
\end{equation}
for the odd parity solution, and
\begin{equation}
\tilde{\rho}(0)=1,\quad \tilde{\rho}'(0)=0
\end{equation}
for the even one.
Next, we give the definition of a resonant or quasilocalized KK mode with the $P-m$ curve defined in Eq. (\ref{P}), which is usually solved by numerical method.
If there is one or more peaks in the $P-m$ curve, then those peaks having full width at half maximum are called resonant peaks and the corresponding KK modes is defined as resonant KK modes. We explain this definition more explicitly. For the $n$-th peak in the $P-m$ curve located at $m=m_n$, whose value is denoted as $P(m_n)$, there must exist two minima around the peak, one at the left hand side of the peak (with $m=m_{n}^{_{-}}$) and another at the right hand side (with $m=m_{n}^{_{+}}$), denoted as $P(m_{n}^{_{-}})$ and $P(m_{n}^{_{+}})$, respectively. If the half value of the $n$-th peak is larger than both  $P(m_{n}^{_{-}})$ and $P(m_{n}^{_{+}})$, i.e., $P(m_n)/2 > P(m_{n}^{_{\pm}})$, then this peak has full width at half maximum and the corresponding massive KK mode with mass $m=m_n$ is a resonant KK mode. The full width at half maximum $\Gamma_n\equiv\Delta m_n$ is defined as the decay width of the $n$-th resonant KK mode and $\tau_n\equiv 1/\Gamma_n$ is defined as its lifetime.
So we can use the $P-m$ curve to check whether there are resonant modes or not in the spectrum of the vector KK modes. Note that, if a peak has no full width at half maximum, then we can not give the lifetime for the corresponding KK mode. Such a peak is not a resonance according to our definition.
}

It is worth to note that in particle physics, the resonance is defined as a peak located around a certain energy found in the cross section which is a function of the total energy of colliding particles. 
For example, for a resonant scattering from an initial two-body state $n$ to a final two-body state $n'$, the corresponding cross-section reads 
\cite{[][{, p. 163.}]Weinberg1995}
\begin{equation}
\sigma(n\to n',E)\propto \frac{\Gamma_{n}\Gamma_{n'}}{(E-E_{R})^{2}+\Gamma^{2}/4},\label{c-s}
\end{equation}
where $E_{R}$ is the energy of the resonance, $\Gamma_{n'}$ is the probability for the resonance decay into the final two-body state $n'$, and $\Gamma$ is the total decay rate which is the sum of all $\Gamma_{n'}$.
The lifetime of a resonance is given by $\tau\equiv 1/\Gamma$ according to the uncertainty principle, where $\Gamma$ is the width of the peak at the half maximum. After some time (lifetime) the resonant particle will decay into more stable particles. While in our definition, the resonances are some mass states for 4D KK particles, which are governed by the Schr\"odinger-like equation \eqref{eq3}.  
The lifetime of a resonance is the time for a 4D KK particle living on branes. Because a resonant KK mode is not localizable, the corresponding KK mode particle will spread into the extra-dimension after some time (lifetime).
The function $P(m)$ in this paper has a similar status with the cross-section function $\sigma(E)$ in the scattering theory, and the same for $m$ and $E$.  Corresponding to the cross-section $\sigma$ in equation \eqref{c-s}, $P(m)$ does not have an analytical form in our paper, but it is effective for us to find out the resonant KK modes.

\begin{figure*}
\begin{center}
\includegraphics[width=0.49\textwidth]{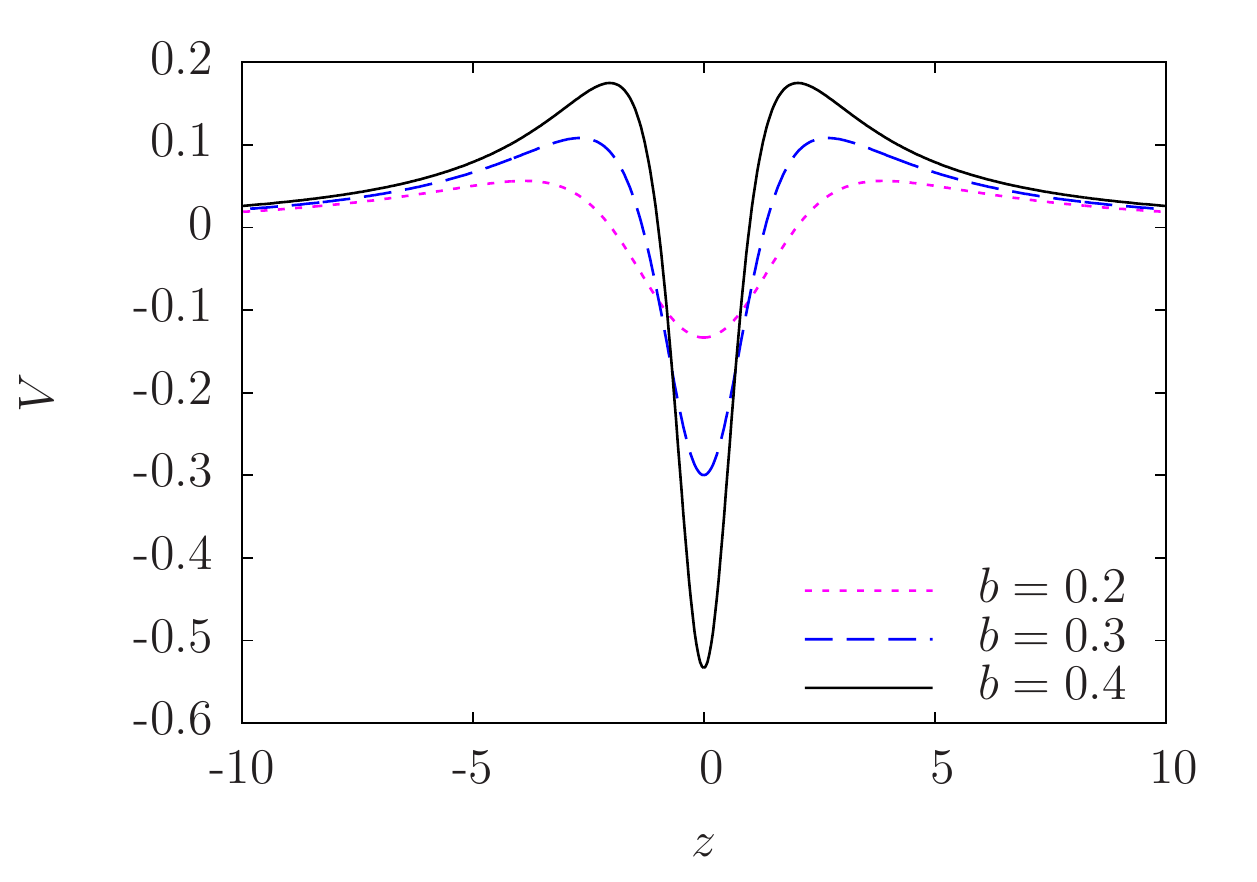}
\includegraphics[width=0.49\textwidth]{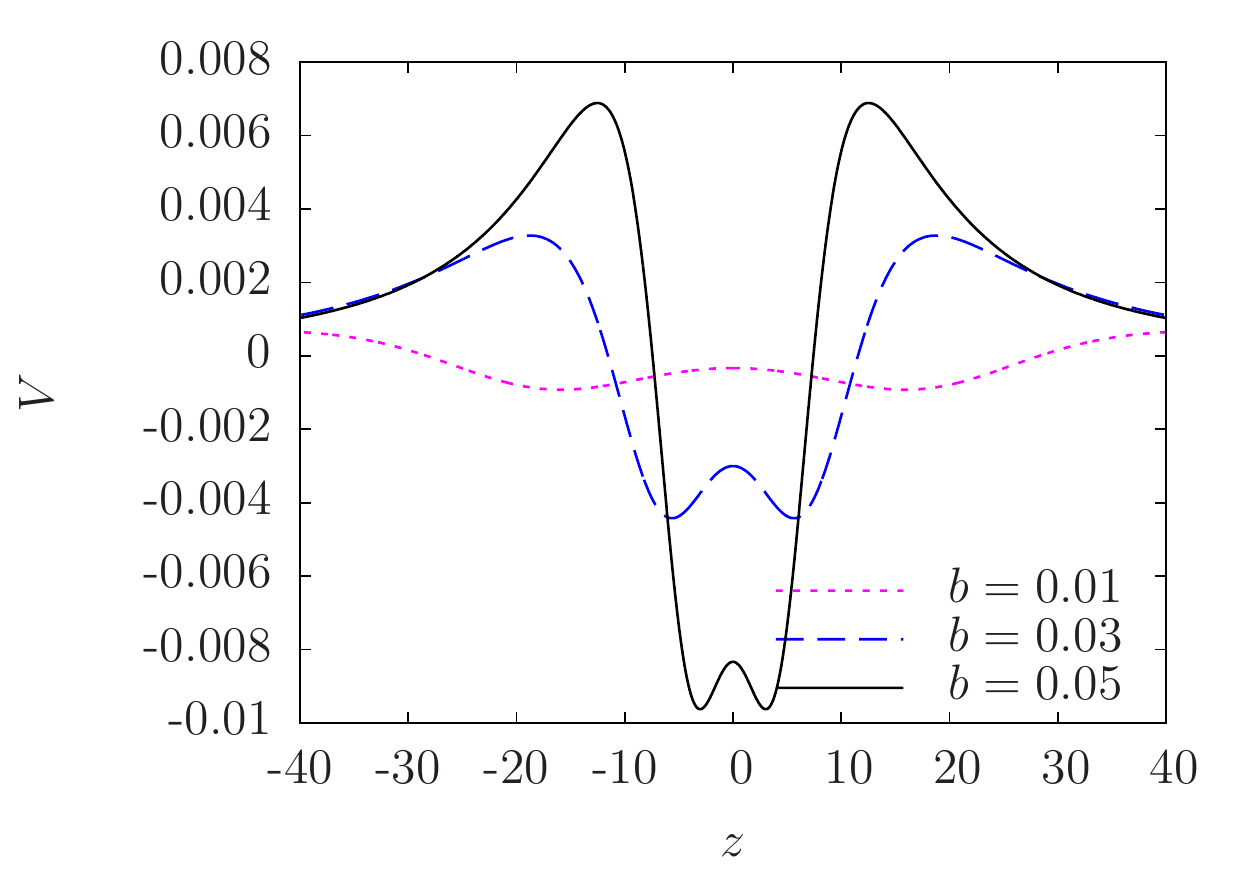}
\caption{The shapes of the effective potential $V(z)$ for the original Bloch brane solution with different values of the parameter $b$.\label{Vz}}
\end{center}
\end{figure*}

\begin{figure*}
\begin{center}
\includegraphics[width=0.49\textwidth]{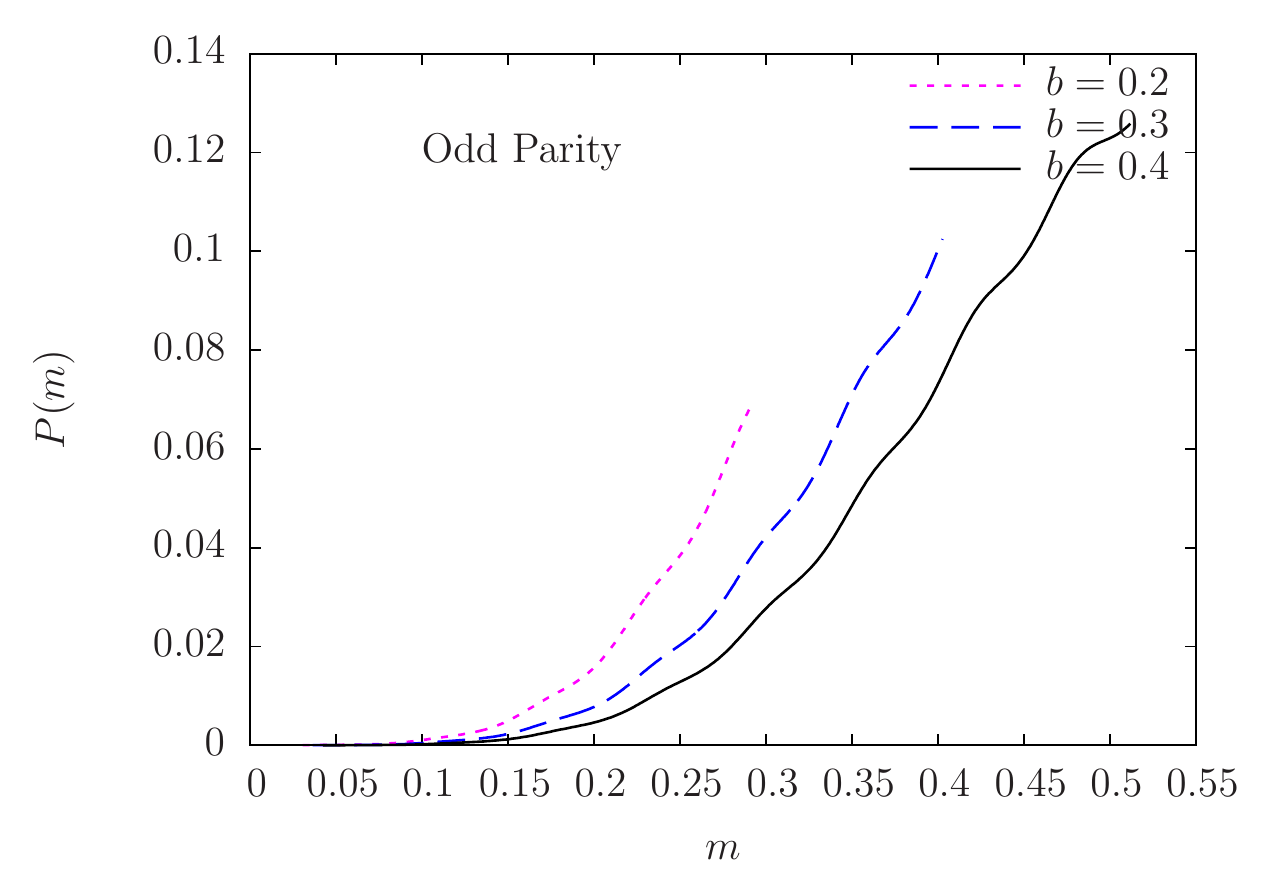}
\includegraphics[width=0.49\textwidth]{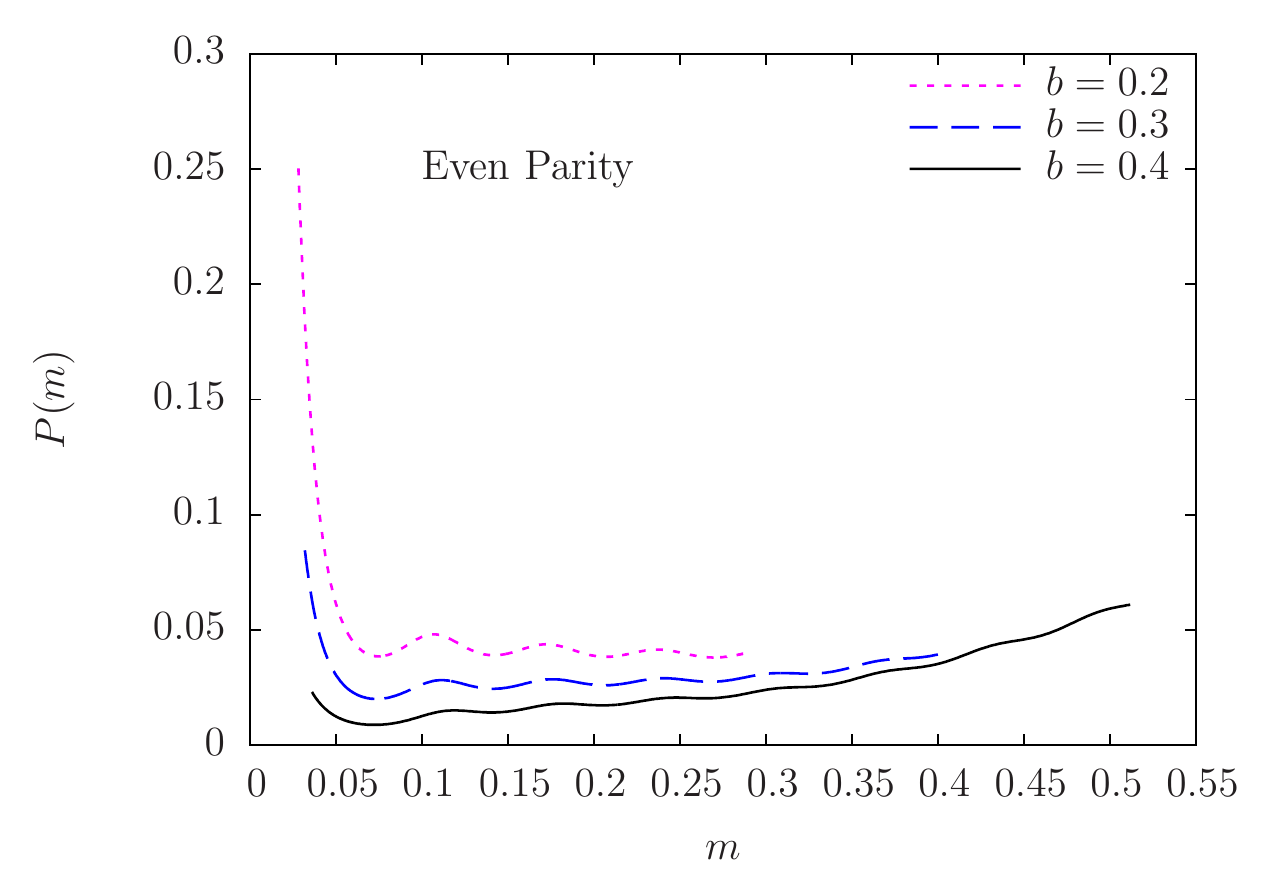}\\
\includegraphics[width=0.49\textwidth]{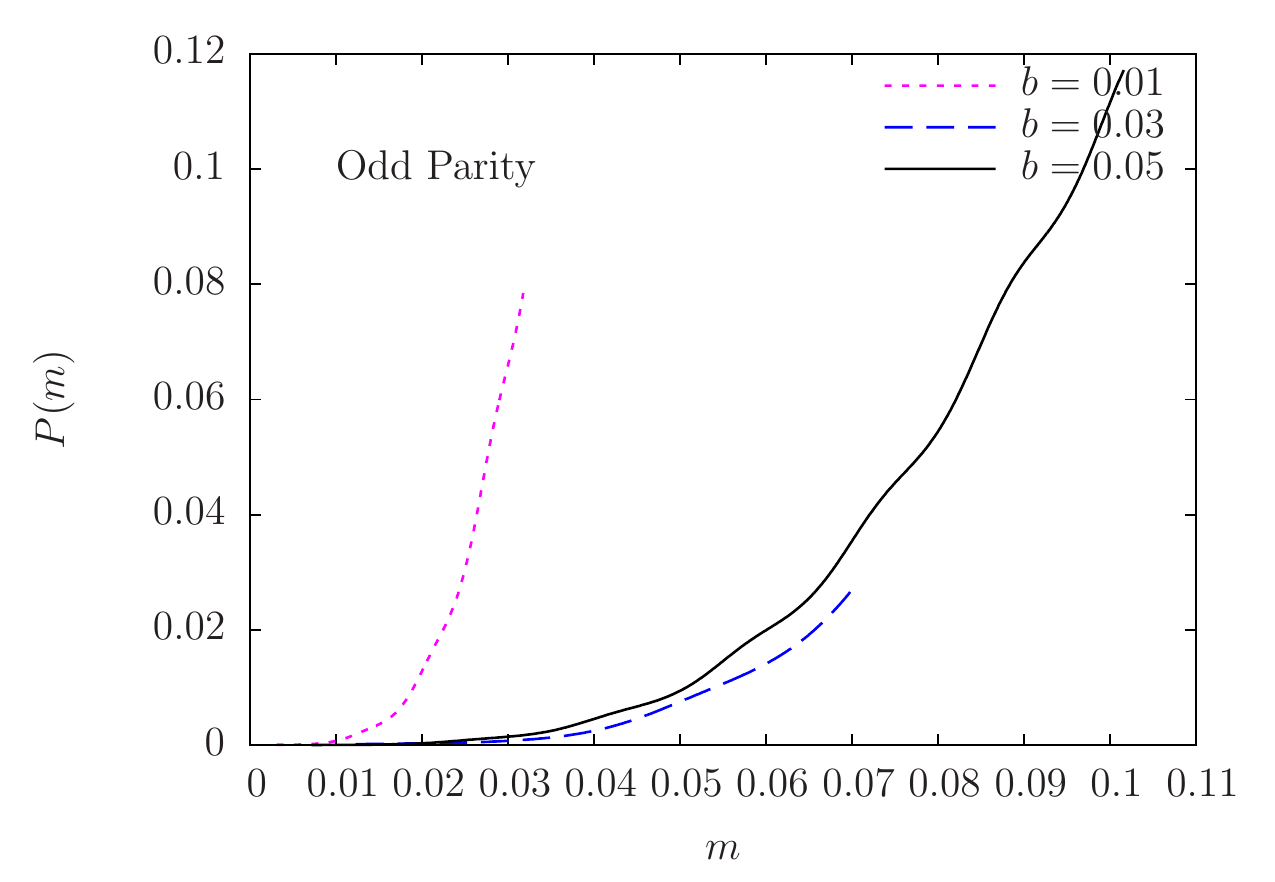}
\includegraphics[width=0.49\textwidth]{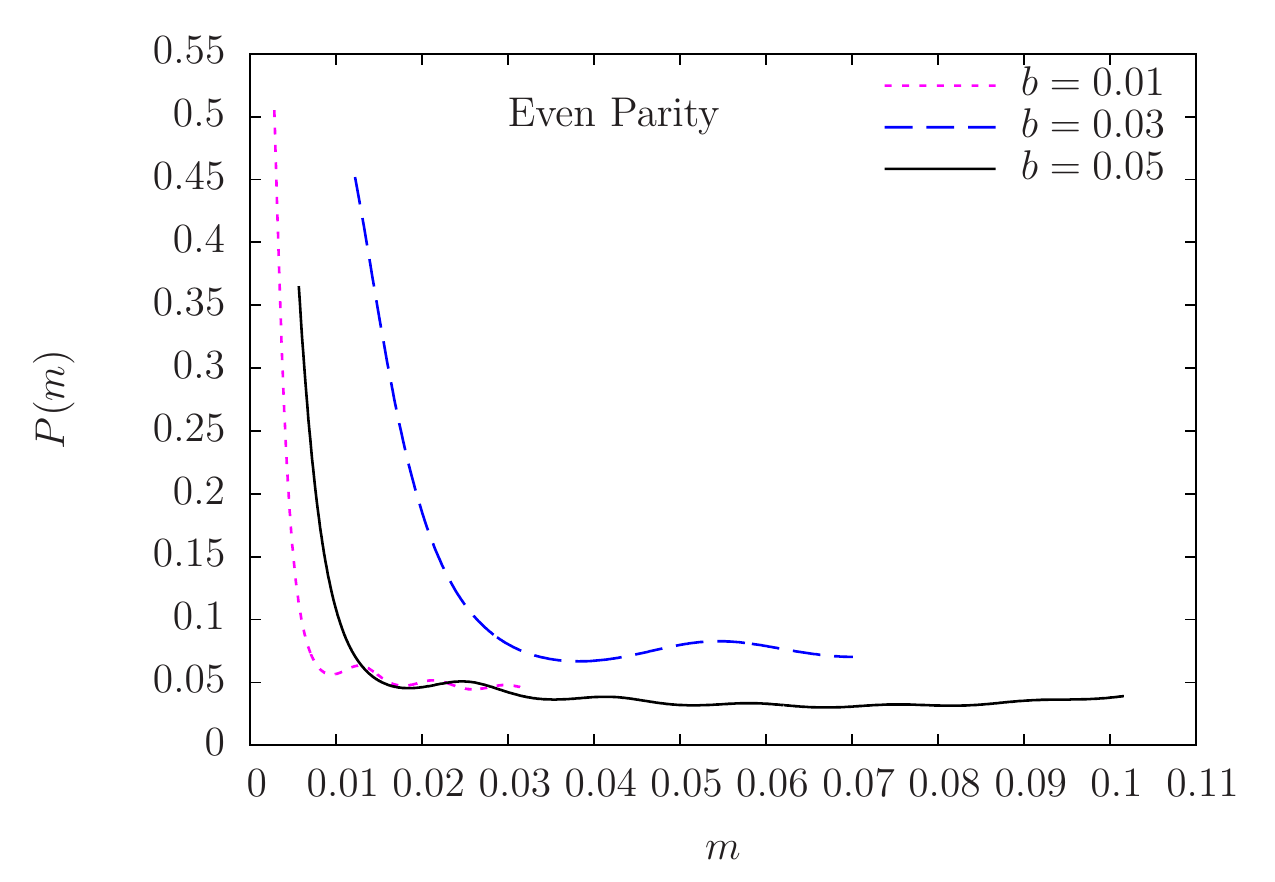}
\caption{The shapes of $P(m)$ as a function of $m$ for different values of the parameter $b$ for the original Bloch brane solution.\label{P1}}
\end{center}
\end{figure*}

 {
For the original Bloch brane solution, the shapes of the effective potential $V(z)$ for different values of the parameter $b$ are shown in Fig.~\ref{Vz}, and the corresponding $P-m$ curves are plotted in Fig.~\ref{P1}. The results are similar for the generalized Bloch solution and we do not show them.
A large range of values of the parameters are checked for both the original and generalized Bloch solutions, but no resonant mode is found.}

\begin{figure*}
\begin{center}
\subfigure[The degenerate Bloch brane solution I]{\label{figDBUz1}
\includegraphics[width=0.45\textwidth]{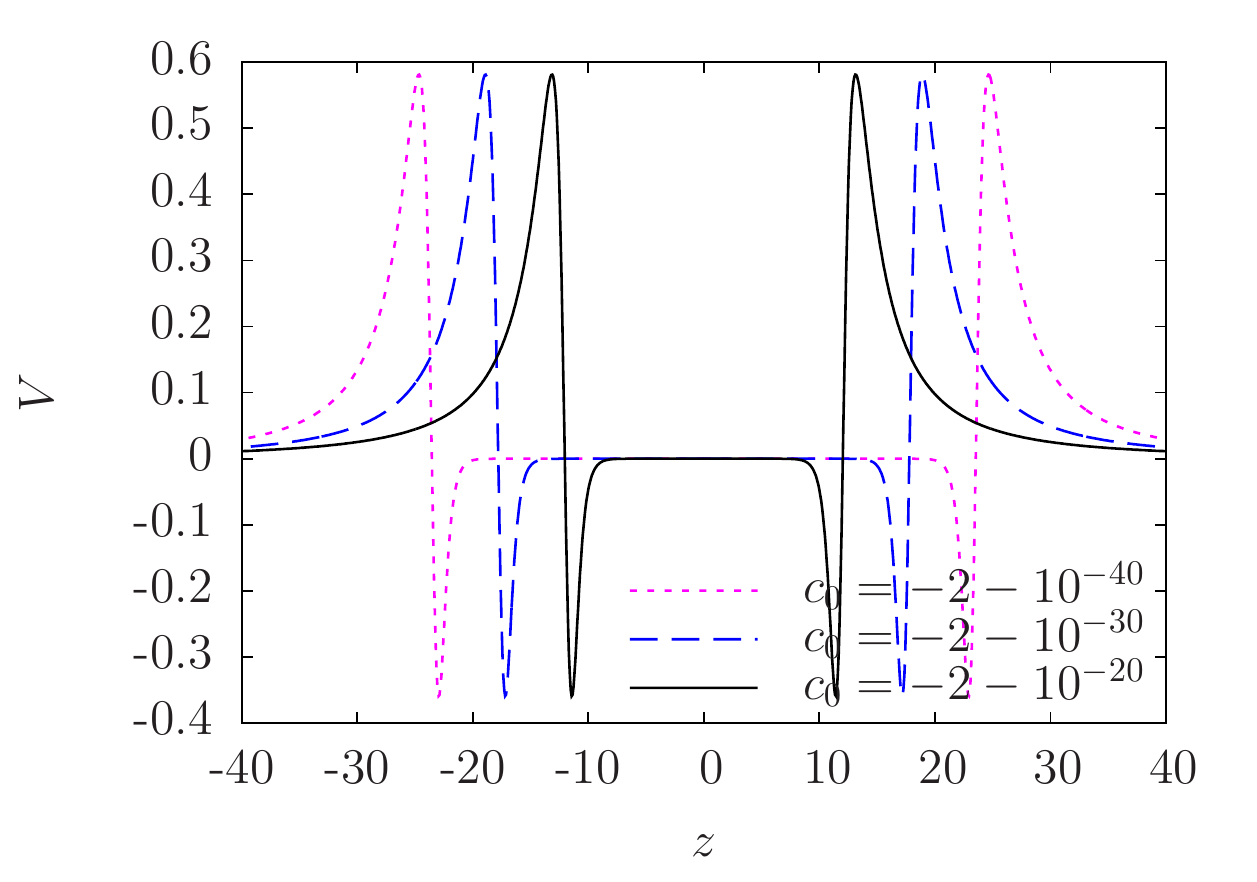}}
\subfigure[The degenerate Bloch brane solution II]{\label{figDBUz2}
\includegraphics[width=0.45\textwidth]{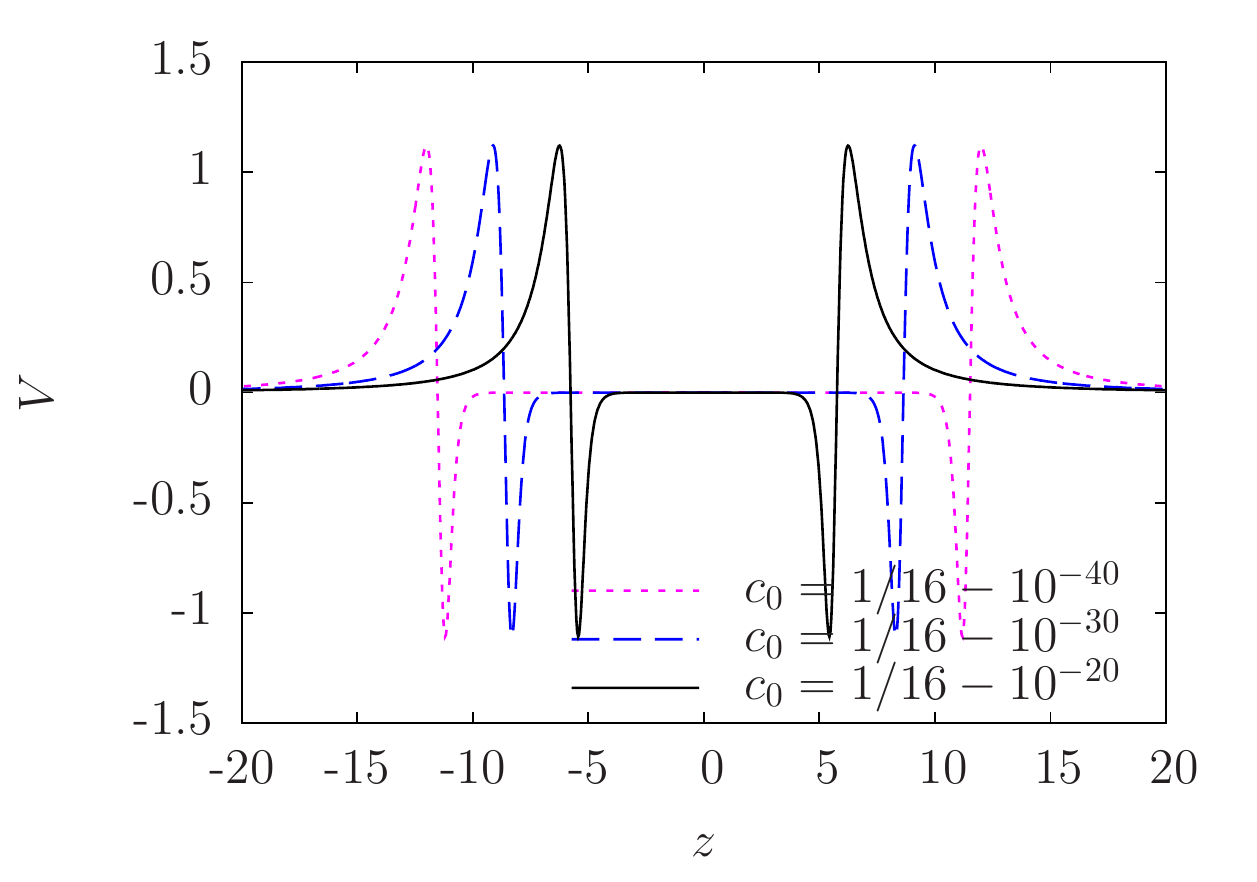}}
\caption{The shapes of the effective potential $V(z)$ with different values of the parameter $c_0$ for the degenerate Bloch brane solutions I (\ref{BlochSolution3}) (left) and II (\ref{BlochSolution4}) (right). The other parameters are set to  $v=1$ and $b=1$.
\label{Vz33}}
\end{center}
\end{figure*}

For the degenerate Bloch brane solutions I and II,
the shapes of $V(z)$ are shown in Fig.~\ref{Vz33}, which shows that the width of the potential well increases with the parameter $d$ and there are two potential wells and two barriers with vanishing potential between them when $d$ is large enough or $c_0\rightarrow -1$ and $c_0\rightarrow1/16$ for solutions I and II, respectively. Here, the parameter $d$ is related to $c_0$ by $c_0=-2-10^{-d}$ and $c_0=1/16-10^{-d}$ for for solutions I and II, respectively.

\begin{figure*}
\begin{center}
\subfigure[$d=10$]{\label{figDBP1}
\includegraphics[width=0.45\textwidth]{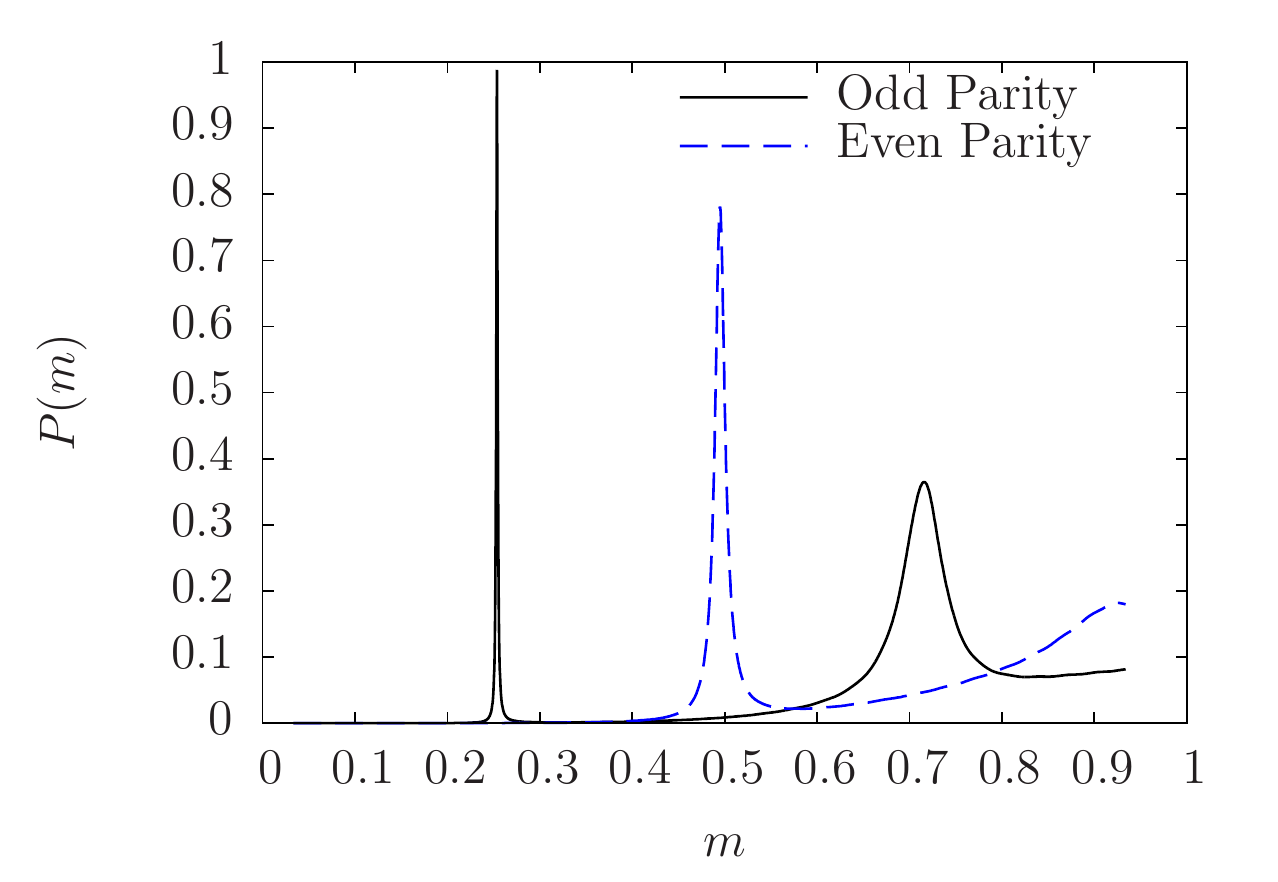}}
\subfigure[$d=20$]{\label{figDBP2}
\includegraphics[width=0.45\textwidth]{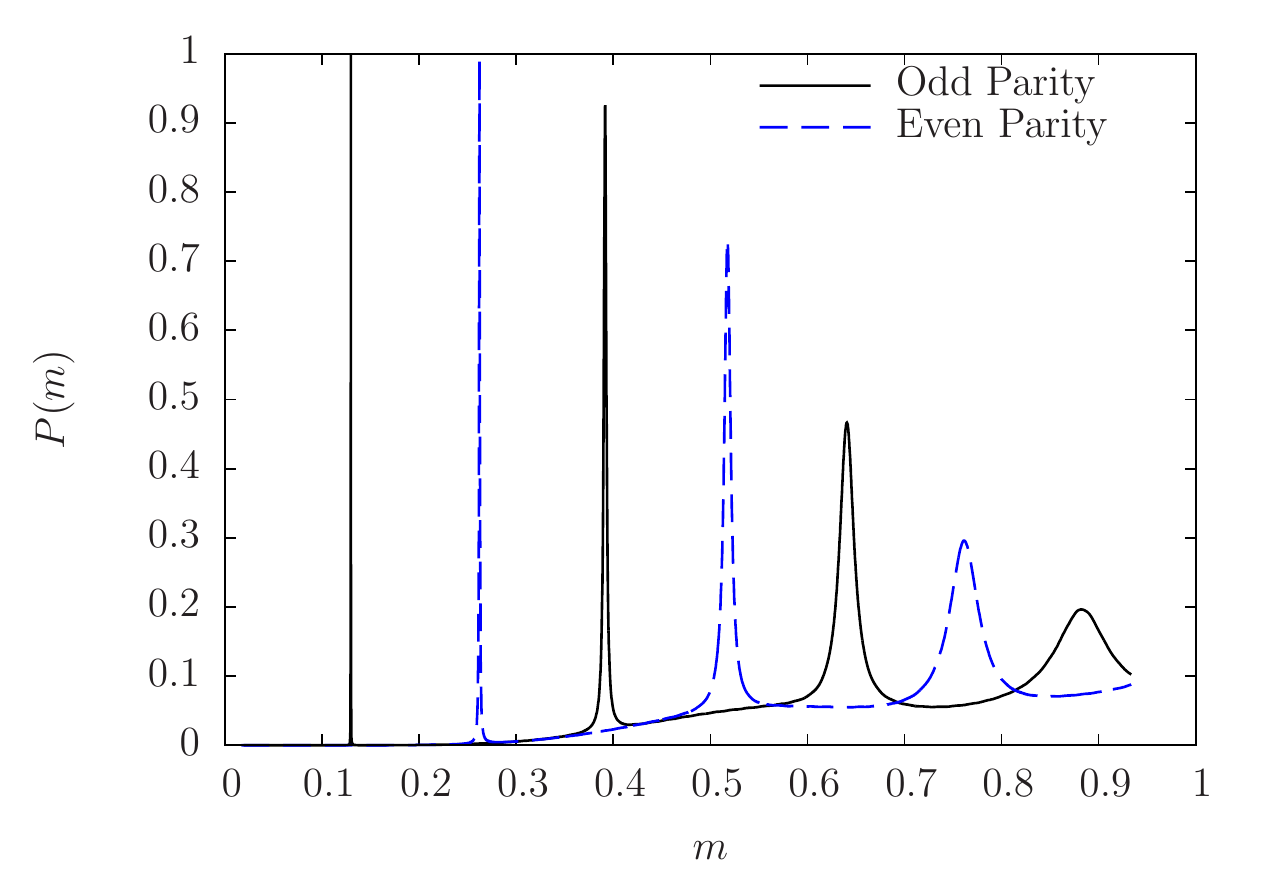}}
\caption{The $P-m$ curves for the degenerate Bloch brane solution I (\ref{BlochSolution3}). The parameters are set to $b=1$, $v=1$, $c_0=-2-10^{-10}$ ($d=10$) (Left), and $c_0=-2-10^{-20}$ ($d=20$) (Right).  \label{DBP1}}
\end{center}
\end{figure*}

\begin{figure*}
\begin{center}
\subfigure[$d=10$]{\label{figDBP21}
\includegraphics[width=0.45\textwidth]{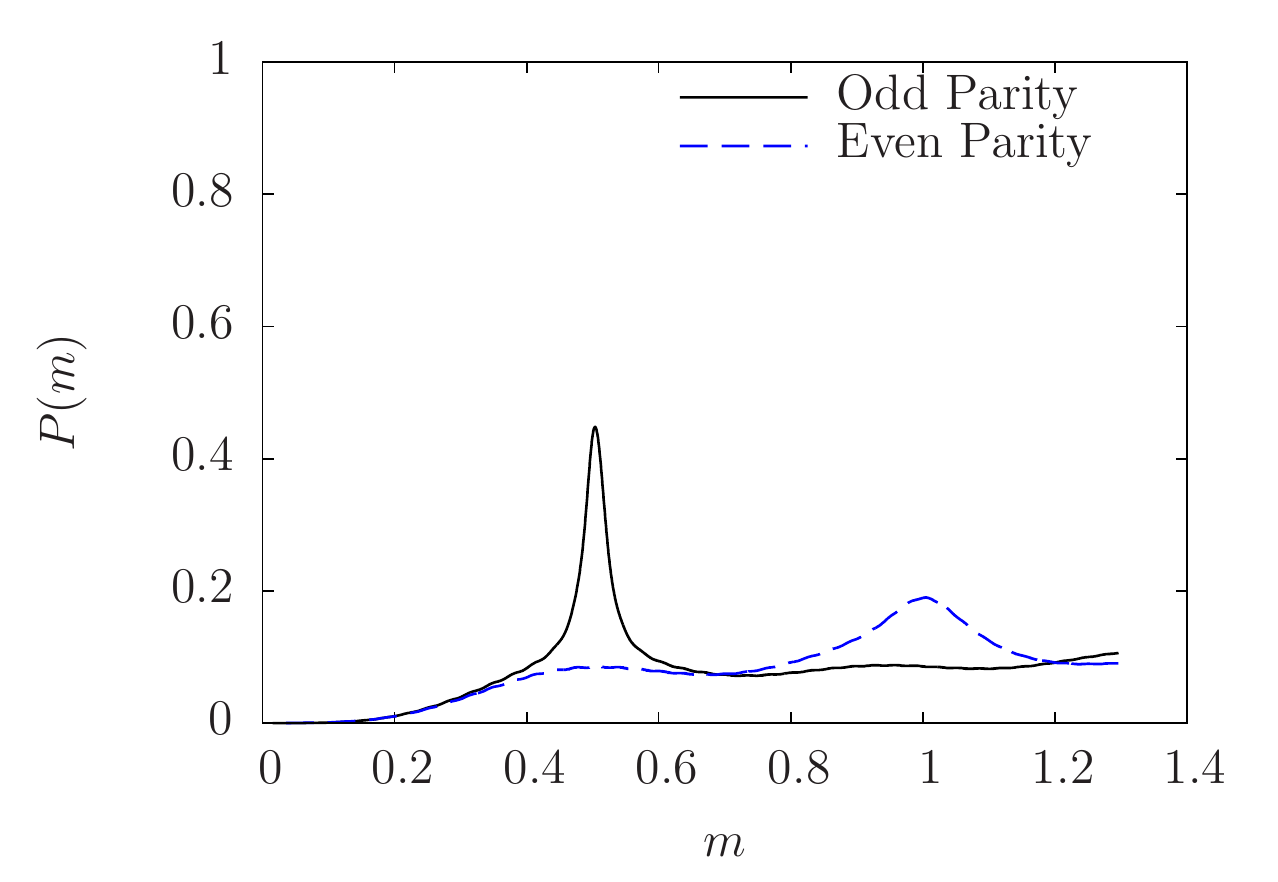}}
\subfigure[$d=20$]{\label{figDBP22}
\includegraphics[width=0.45\textwidth]{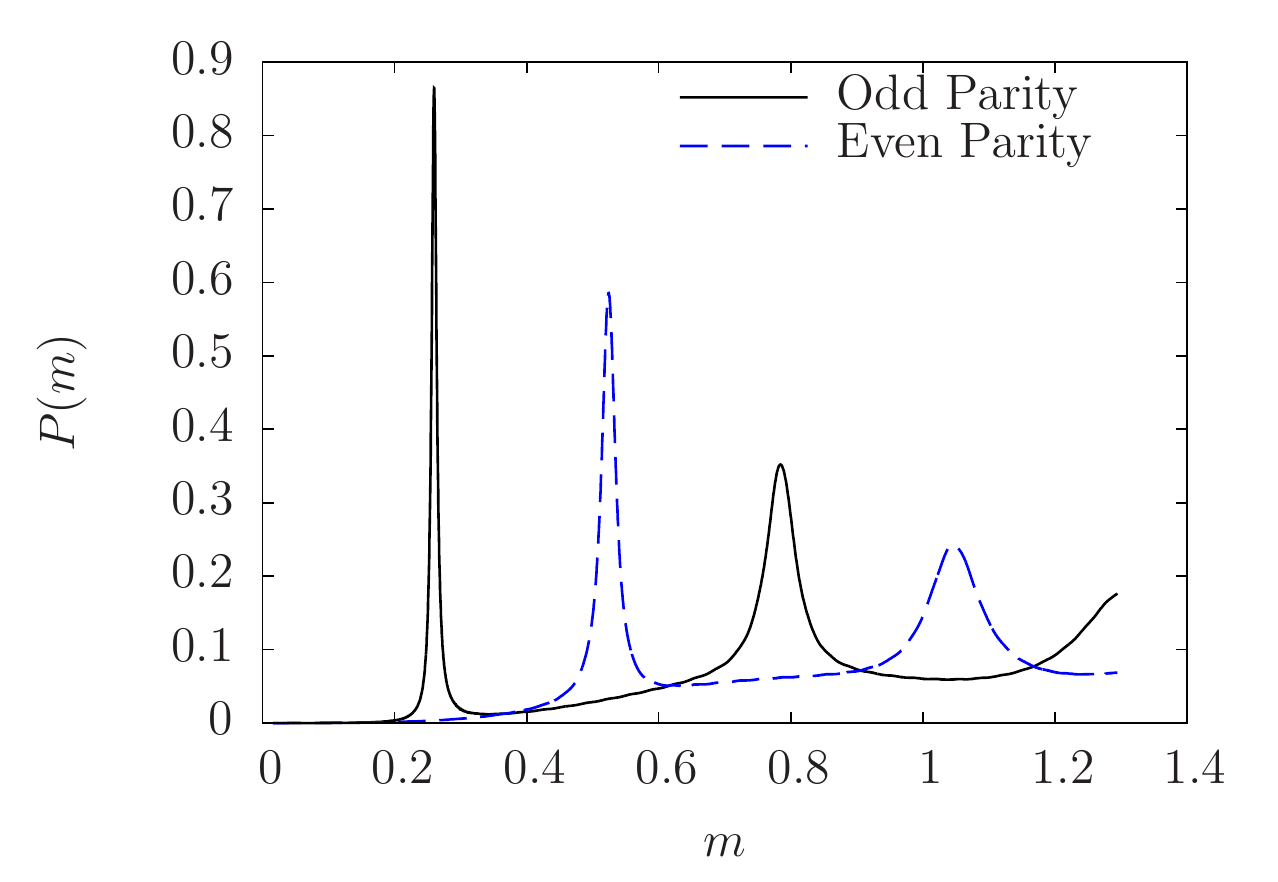}}
\caption{The $P-m$ curves for the degenerate Bloch brane solution II (\ref{BlochSolution4}). The parameters are set to $b=1$, $v=1$, $c_0=1/16-10^{-10}$ ($d=10$) (Left), and $c_0=1/16-10^{-20}$ ($d=20$) (Right).  \label{DBP2}}
\end{center}
\end{figure*}

\begin{figure*}
\begin{center}
\includegraphics[width=0.485\textwidth]{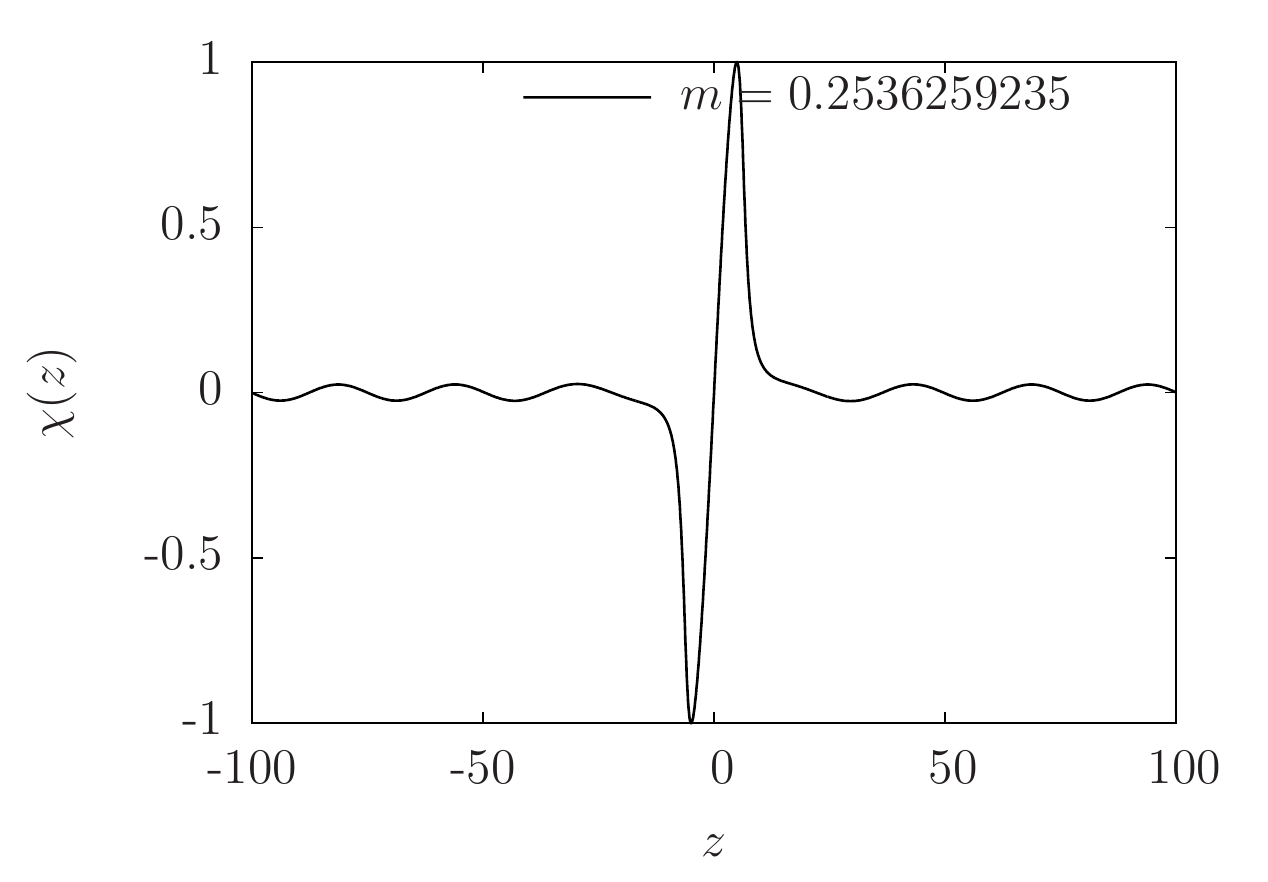}
\includegraphics[width=0.485\textwidth]{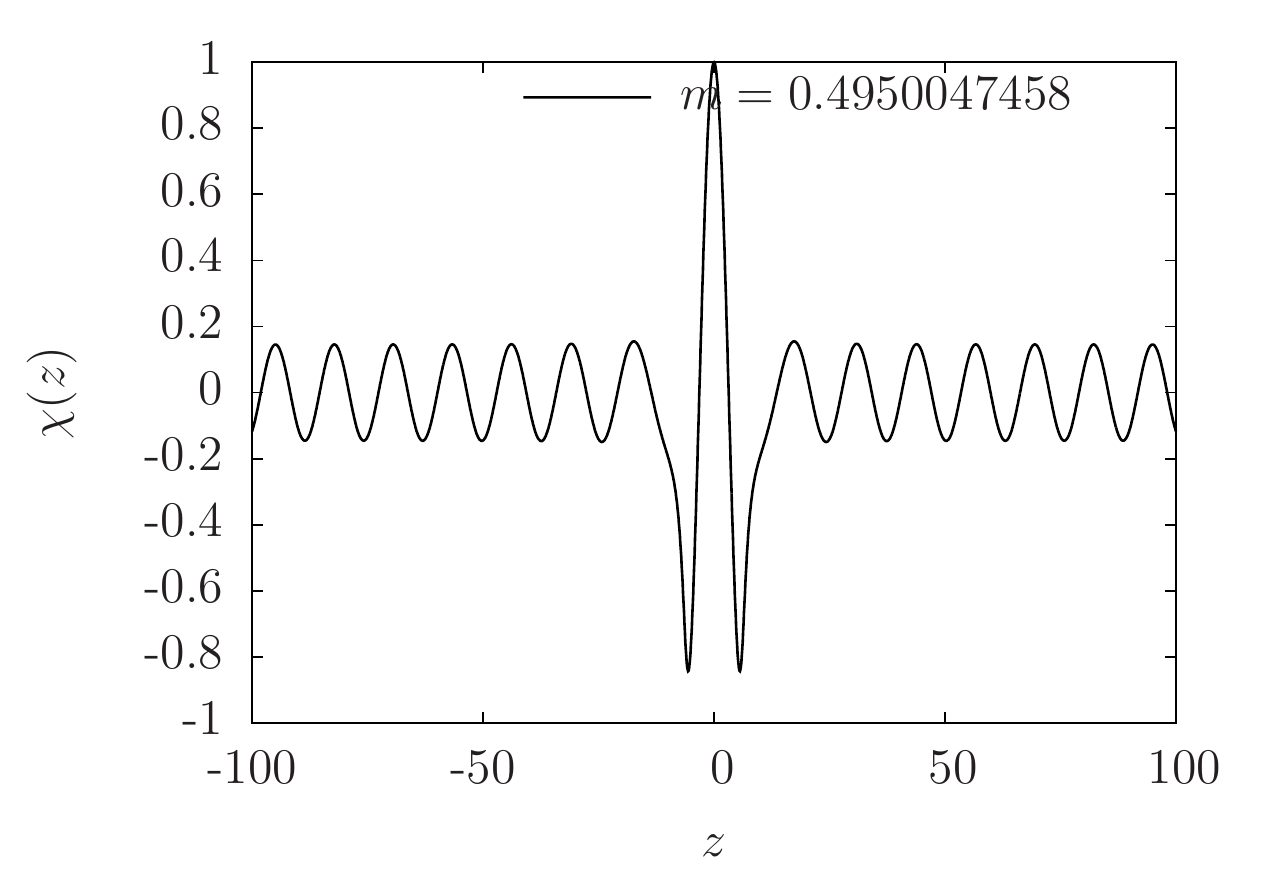}
\includegraphics[width=0.485\textwidth]{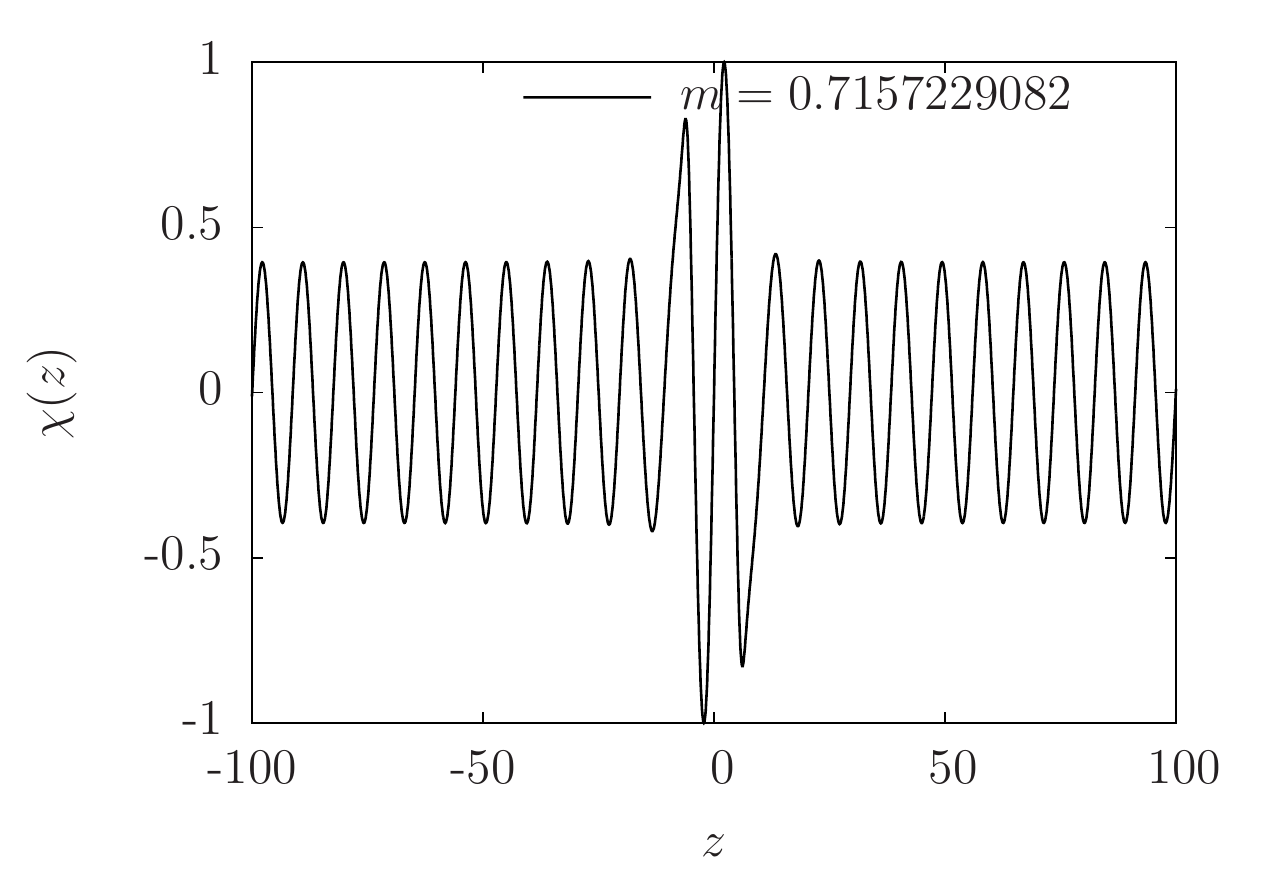}
\caption{The shapes of the resonant modes corresponding to the three peaks in Fig.~\ref{figDBP1}.}\label{resmode}
\end{center}
\end{figure*}

With the degenerate Bloch brane solutions I and II, we find  resonances. In oder to show the result intuitively, we plotted some curves of $P(m)$ in Figs.~\ref{DBP1} and \ref{DBP2} for the degenerate Bloch brane solutions I and  II, respectively. In the curves of $P(m)$, every peak corresponds to a resonant mode.
And by comparing Fig.~\ref{DBP1} and Fig. \ref{DBP2}, we find that the number of the resonant modes increases with the width of the degenerate Bloch branes. The shapes of the resonant modes corresponding to the three peaks in Fig.~\ref{figDBP1} are shown in Fig.~\ref{resmode}, from which it can be seen that the resonant KK modes with larger mass and shorter lifetime are nearly plane waves, while the resonances with lower mass and longer lifetime are quasibound modes.

\section{Conclusions}\label{Cons}

We have studied the localization of U(1) gauge field on the Bloch brane with the CHH mechanism.
 {There are two scalar fields in the Bloch brane model. In the CHH mechanism, one of scalar fields couples directly to the U(1) gauge field.} So, compared to the KT mechanism, the CHH mechanism is simpler to study the localization of $U(1)$ gauge field in the Bloch brane scenario.

In this work, four kinds of Bloch brane solutions were discussed, they are the original, generalized, and degenerate I and II Bloch brane solutions, respectively.  {
With the CHH mechanism, the Schr\"odinger-like equation for the vector KK modes can be recast to the supersymmetric quantum mechanics form,}
so the tachyonic KK modes are excluded.
We found that the zero mode of the U(1) gauge field can be localized on the brane and the mass spectrum is continuous with $m^2\geq0$. The resonant modes in the KK spectrum were also discussed. For the original and generalized Bloch brane solutions, we did not find any resonant mode. While for the degenerate Bloch brane I and II solutions, we found some resonant modes, and the number of resonant modes is related with the inner structure of the Bloch brane and increases with the brane width.

\section*{Acknowledgments}

We are grateful to the referee for her/his comments and suggestions, which are very useful for us to improve this manuscript.
Z-H Zhao is supported by the National Natural Science Foundation of China (Grant No.  11305095), the Natural Science Foundation of Shandong Province, China (Grant No. ZR2013AQ016), and Scientific Research Foundation of Shandong University of Science and Technology for Recruited Talents (Grant No. 2013RCJJ026).
Y-X Liu is supported by
the National Natural Science Foundation of China (Grant No. 11375075), and the Fundamental Research Funds for the Central Universities (Grant No. lzujbky-2013-18).
Y. Zhong is supported by the scholarship granted by the Chinese Scholarship Council (CSC).

\bibliography{/Users/zhaozhenhua/yunpan/library_zhao/articles_all}

\begin{thebibliography}{44}%
\makeatletter
\providecommand \@ifxundefined [1]{%
 \@ifx{#1\undefined}
}%
\providecommand \@ifnum [1]{%
 \ifnum #1\expandafter \@firstoftwo
 \else \expandafter \@secondoftwo
 \fi
}%
\providecommand \@ifx [1]{%
 \ifx #1\expandafter \@firstoftwo
 \else \expandafter \@secondoftwo
 \fi
}%
\providecommand \natexlab [1]{#1}%
\providecommand \enquote  [1]{``#1''}%
\providecommand \bibnamefont  [1]{#1}%
\providecommand \bibfnamefont [1]{#1}%
\providecommand \citenamefont [1]{#1}%
\providecommand \href@noop [0]{\@secondoftwo}%
\providecommand \href [0]{\begingroup \@sanitize@url \@href}%
\providecommand \@href[1]{\@@startlink{#1}\@@href}%
\providecommand \@@href[1]{\endgroup#1\@@endlink}%
\providecommand \@sanitize@url [0]{\catcode `\\12\catcode `\$12\catcode
  `\&12\catcode `\#12\catcode `\^12\catcode `\_12\catcode `\%12\relax}%
\providecommand \@@startlink[1]{}%
\providecommand \@@endlink[0]{}%
\providecommand \url  [0]{\begingroup\@sanitize@url \@url }%
\providecommand \@url [1]{\endgroup\@href {#1}{\urlprefix }}%
\providecommand \urlprefix  [0]{URL }%
\providecommand \Eprint [0]{\href }%
\providecommand \doibase [0]{http://dx.doi.org/}%
\providecommand \selectlanguage [0]{\@gobble}%
\providecommand \bibinfo  [0]{\@secondoftwo}%
\providecommand \bibfield  [0]{\@secondoftwo}%
\providecommand \translation [1]{[#1]}%
\providecommand \BibitemOpen [0]{}%
\providecommand \bibitemStop [0]{}%
\providecommand \bibitemNoStop [0]{.\EOS\space}%
\providecommand \EOS [0]{\spacefactor3000\relax}%
\providecommand \BibitemShut  [1]{\csname bibitem#1\endcsname}%
\let\auto@bib@innerbib\@empty
\bibitem [{\citenamefont {Randall}\ and\ \citenamefont
  {Sundrum}(1999{\natexlab{a}})}]{Randall199983}%
  \BibitemOpen
  \bibfield  {author} {\bibinfo {author} {\bibfnamefont {L.}~\bibnamefont
  {Randall}}\ and\ \bibinfo {author} {\bibfnamefont {R.}~\bibnamefont
  {Sundrum}},\ }\href {\doibase 10.1103/PhysRevLett.83.3370} {\bibfield
  {journal} {\bibinfo  {journal} {Phys. Rev. Lett.}\ }\textbf {\bibinfo
  {volume} {83}},\ \bibinfo {pages} {3370} (\bibinfo {year}
  {1999}{\natexlab{a}})},\ \Eprint {http://arxiv.org/abs/hep-ph/9905221}
  {arXiv:hep-ph/9905221} \BibitemShut {NoStop}%
\bibitem [{\citenamefont {Randall}\ and\ \citenamefont
  {Sundrum}(1999{\natexlab{b}})}]{Randall199983a}%
  \BibitemOpen
  \bibfield  {author} {\bibinfo {author} {\bibfnamefont {L.}~\bibnamefont
  {Randall}}\ and\ \bibinfo {author} {\bibfnamefont {R.}~\bibnamefont
  {Sundrum}},\ }\href {\doibase 10.1103/PhysRevLett.83.4690} {\bibfield
  {journal} {\bibinfo  {journal} {Phys. Rev. Lett.}\ }\textbf {\bibinfo
  {volume} {83}},\ \bibinfo {pages} {4690} (\bibinfo {year}
  {1999}{\natexlab{b}})},\ \Eprint {http://arxiv.org/abs/hep-th/9906064}
  {arXiv:hep-th/9906064} \BibitemShut {NoStop}%
\bibitem [{\citenamefont {Gremm}(2000)}]{Gremm2000478}%
  \BibitemOpen
  \bibfield  {author} {\bibinfo {author} {\bibfnamefont {M.}~\bibnamefont
  {Gremm}},\ }\href {\doibase 10.1016/S0370-2693(00)00303-8} {\bibfield
  {journal} {\bibinfo  {journal} {Phys. Lett. B}\ }\textbf {\bibinfo {volume}
  {478}},\ \bibinfo {pages} {434} (\bibinfo {year} {2000})},\ \Eprint
  {http://arxiv.org/abs/hep-th/9912060} {arXiv:hep-th/9912060} \BibitemShut
  {NoStop}%
\bibitem [{\citenamefont {Rubakov}\ and\ \citenamefont
  {Shaposhnikov}(1983)}]{Rubakov1983}%
  \BibitemOpen
  \bibfield  {author} {\bibinfo {author} {\bibfnamefont {V.~A.}\ \bibnamefont
  {Rubakov}}\ and\ \bibinfo {author} {\bibfnamefont {M.~E.}\ \bibnamefont
  {Shaposhnikov}},\ }\href {\doibase 10.1016/0370-2693(83)91253-4} {\bibfield
  {journal} {\bibinfo  {journal} {Phys. Lett. B}\ }\textbf {\bibinfo {volume}
  {125}},\ \bibinfo {pages} {136} (\bibinfo {year} {1983})}\BibitemShut
  {NoStop}%
\bibitem [{\citenamefont {Bajc}\ and\ \citenamefont
  {Gabadadze}(2000)}]{Bajc2000}%
  \BibitemOpen
  \bibfield  {author} {\bibinfo {author} {\bibfnamefont {B.}~\bibnamefont
  {Bajc}}\ and\ \bibinfo {author} {\bibfnamefont {G.}~\bibnamefont
  {Gabadadze}},\ }\href {\doibase 10.1016/S0370-2693(00)00055-1} {\bibfield
  {journal} {\bibinfo  {journal} {Phys. Lett. B}\ }\textbf {\bibinfo {volume}
  {474}},\ \bibinfo {pages} {282} (\bibinfo {year} {2000})},\ \Eprint
  {http://arxiv.org/abs/hep-th/9912232} {arXiv:hep-th/9912232} \BibitemShut
  {NoStop}%
\bibitem [{\citenamefont {Randjbar-Daemi}\ and\ \citenamefont
  {Shaposhnikov}(2000)}]{Randjbar-Daemi2000}%
  \BibitemOpen
  \bibfield  {author} {\bibinfo {author} {\bibfnamefont {S.}~\bibnamefont
  {Randjbar-Daemi}}\ and\ \bibinfo {author} {\bibfnamefont {M.~E.}\
  \bibnamefont {Shaposhnikov}},\ }\href {\doibase
  10.1016/S0370-2693(00)01100-X} {\bibfield  {journal} {\bibinfo  {journal}
  {Phys. Lett. B}\ }\textbf {\bibinfo {volume} {492}},\ \bibinfo {pages} {361}
  (\bibinfo {year} {2000})},\ \Eprint {http://arxiv.org/abs/hep-th/0008079}
  {arXiv:hep-th/0008079} \BibitemShut {NoStop}%
\bibitem [{\citenamefont {Ringeval}\ \emph {et~al.}(2002)\citenamefont
  {Ringeval}, \citenamefont {Peter},\ and\ \citenamefont
  {Uzan}}]{Ringeval200265}%
  \BibitemOpen
  \bibfield  {author} {\bibinfo {author} {\bibfnamefont {C.}~\bibnamefont
  {Ringeval}}, \bibinfo {author} {\bibfnamefont {P.}~\bibnamefont {Peter}}, \
  and\ \bibinfo {author} {\bibfnamefont {J.-P.}\ \bibnamefont {Uzan}},\ }\href
  {\doibase 10.1103/PhysRevD.65.044016} {\bibfield  {journal} {\bibinfo
  {journal} {Phys. Rev. D}\ }\textbf {\bibinfo {volume} {65}},\ \bibinfo
  {pages} {044016} (\bibinfo {year} {2002})},\ \Eprint
  {http://arxiv.org/abs/hep-th/0109194} {arXiv:hep-th/0109194} \BibitemShut
  {NoStop}%
\bibitem [{\citenamefont {Koley}\ and\ \citenamefont
  {Kar}(2005)}]{Koley200522}%
  \BibitemOpen
  \bibfield  {author} {\bibinfo {author} {\bibfnamefont {R.}~\bibnamefont
  {Koley}}\ and\ \bibinfo {author} {\bibfnamefont {S.}~\bibnamefont {Kar}},\
  }\href {\doibase 10.1088/0264-9381/22/4/008} {\bibfield  {journal} {\bibinfo
  {journal} {Class. Quant. Grav.}\ }\textbf {\bibinfo {volume} {22}},\ \bibinfo
  {pages} {753} (\bibinfo {year} {2005})},\ \Eprint
  {http://arxiv.org/abs/hep-th/0407158} {arXiv:hep-th/0407158} \BibitemShut
  {NoStop}%
\bibitem [{\citenamefont {Melfo}\ \emph {et~al.}(2006)\citenamefont {Melfo},
  \citenamefont {Pantoja},\ and\ \citenamefont {Tempo}}]{Melfo2006}%
  \BibitemOpen
  \bibfield  {author} {\bibinfo {author} {\bibfnamefont {A.}~\bibnamefont
  {Melfo}}, \bibinfo {author} {\bibfnamefont {N.}~\bibnamefont {Pantoja}}, \
  and\ \bibinfo {author} {\bibfnamefont {J.~D.}\ \bibnamefont {Tempo}},\ }\href
  {\doibase 10.1103/PhysRevD.73.044033} {\bibfield  {journal} {\bibinfo
  {journal} {Phys. Rev. D}\ }\textbf {\bibinfo {volume} {73}},\ \bibinfo
  {pages} {044033} (\bibinfo {year} {2006})},\ \Eprint
  {http://arxiv.org/abs/hep-th/0601161} {arXiv:hep-th/0601161} \BibitemShut
  {NoStop}%
\bibitem [{\citenamefont {Liu}\ \emph {et~al.}(2008{\natexlab{a}})\citenamefont
  {Liu}, \citenamefont {Zhang}, \citenamefont {Zhang},\ and\ \citenamefont
  {Duan}}]{LiuZhangZhangDuan2008}%
  \BibitemOpen
  \bibfield  {author} {\bibinfo {author} {\bibfnamefont {Y.-X.}\ \bibnamefont
  {Liu}}, \bibinfo {author} {\bibfnamefont {L.-D.}\ \bibnamefont {Zhang}},
  \bibinfo {author} {\bibfnamefont {L.-J.}\ \bibnamefont {Zhang}}, \ and\
  \bibinfo {author} {\bibfnamefont {Y.-S.}\ \bibnamefont {Duan}},\ }\href
  {\doibase 10.1103/PhysRevD.78.065025} {\bibfield  {journal} {\bibinfo
  {journal} {Phys. Rev. D}\ }\textbf {\bibinfo {volume} {78}},\ \bibinfo
  {pages} {065025} (\bibinfo {year} {2008}{\natexlab{a}})},\ \Eprint
  {http://arxiv.org/abs/0804.4553} {arXiv:0804.4553 [hep-th]} \BibitemShut
  {NoStop}%
\bibitem [{\citenamefont {Liu}\ \emph {et~al.}(2013)\citenamefont {Liu},
  \citenamefont {Xu}, \citenamefont {Chen},\ and\ \citenamefont
  {Wei}}]{LiuXuChenWei2013}%
  \BibitemOpen
  \bibfield  {author} {\bibinfo {author} {\bibfnamefont {Y.-X.}\ \bibnamefont
  {Liu}}, \bibinfo {author} {\bibfnamefont {Z.-G.}\ \bibnamefont {Xu}},
  \bibinfo {author} {\bibfnamefont {F.-W.}\ \bibnamefont {Chen}}, \ and\
  \bibinfo {author} {\bibfnamefont {S.-W.}\ \bibnamefont {Wei}},\ }\href@noop
  {} {\enquote {\bibinfo {title} {New localization mechanism of fermions on
  braneworlds},}\ } (\bibinfo {year} {2013}),\ \Eprint
  {http://arxiv.org/abs/1312.4145} {arXiv:1312.4145 [hep-th]} \BibitemShut
  {NoStop}%
\bibitem [{\citenamefont {Pomarol}(2000)}]{Pomarol2000}%
  \BibitemOpen
  \bibfield  {author} {\bibinfo {author} {\bibfnamefont {A.}~\bibnamefont
  {Pomarol}},\ }\href {\doibase 10.1016/S0370-2693(00)00737-1} {\bibfield
  {journal} {\bibinfo  {journal} {Phys. Lett. B}\ }\textbf {\bibinfo {volume}
  {486}},\ \bibinfo {pages} {153} (\bibinfo {year} {2000})},\ \Eprint
  {http://arxiv.org/abs/hep-ph/9911294} {arXiv:hep-ph/9911294} \BibitemShut
  {NoStop}%
\bibitem [{\citenamefont {Oda}(2001)}]{Oda2001}%
  \BibitemOpen
  \bibfield  {author} {\bibinfo {author} {\bibfnamefont {I.}~\bibnamefont
  {Oda}},\ }\href@noop {} {\enquote {\bibinfo {title} {A new mechanism for
  trapping of photon},}\ } (\bibinfo {year} {2001}),\ \Eprint
  {http://arxiv.org/abs/hep-th/0103052} {arXiv:hep-th/0103052} \BibitemShut
  {NoStop}%
\bibitem [{\citenamefont {Oda}(2000)}]{Oda2000}%
  \BibitemOpen
  \bibfield  {author} {\bibinfo {author} {\bibfnamefont {I.}~\bibnamefont
  {Oda}},\ }\href {\doibase 10.1016/S0370-2693(00)01284-3} {\bibfield
  {journal} {\bibinfo  {journal} {Phys. Lett. B}\ }\textbf {\bibinfo {volume}
  {496}},\ \bibinfo {pages} {113} (\bibinfo {year} {2000})},\ \Eprint
  {http://arxiv.org/abs/hep-th/0006203} {arXiv:hep-th/0006203 [hep-th]}
  \BibitemShut {NoStop}%
\bibitem [{\citenamefont {Dvali}\ \emph {et~al.}(2001)\citenamefont {Dvali},
  \citenamefont {Gabadadze},\ and\ \citenamefont {Shifman}}]{Dvali2001}%
  \BibitemOpen
  \bibfield  {author} {\bibinfo {author} {\bibfnamefont {G.~R.}\ \bibnamefont
  {Dvali}}, \bibinfo {author} {\bibfnamefont {G.}~\bibnamefont {Gabadadze}}, \
  and\ \bibinfo {author} {\bibfnamefont {M.~A.}\ \bibnamefont {Shifman}},\
  }\href {\doibase 10.1016/S0370-2693(00)01329-0} {\bibfield  {journal}
  {\bibinfo  {journal} {Phys. Lett. B}\ }\textbf {\bibinfo {volume} {497}},\
  \bibinfo {pages} {271} (\bibinfo {year} {2001})},\ \Eprint
  {http://arxiv.org/abs/hep-th/0010071} {arXiv:hep-th/0010071} \BibitemShut
  {NoStop}%
\bibitem [{\citenamefont {Dimopoulos}\ \emph {et~al.}(2001)\citenamefont
  {Dimopoulos}, \citenamefont {Farakos}, \citenamefont {Kehagias},\ and\
  \citenamefont {Koutsoumbas}}]{Dimopoulos2001}%
  \BibitemOpen
  \bibfield  {author} {\bibinfo {author} {\bibfnamefont {P.}~\bibnamefont
  {Dimopoulos}}, \bibinfo {author} {\bibfnamefont {K.}~\bibnamefont {Farakos}},
  \bibinfo {author} {\bibfnamefont {A.}~\bibnamefont {Kehagias}}, \ and\
  \bibinfo {author} {\bibfnamefont {G.}~\bibnamefont {Koutsoumbas}},\ }\href
  {\doibase 10.1016/S0550-3213(01)00451-5} {\bibfield  {journal} {\bibinfo
  {journal} {Nucl. Phys. B}\ }\textbf {\bibinfo {volume} {617}},\ \bibinfo
  {pages} {237} (\bibinfo {year} {2001})},\ \Eprint
  {http://arxiv.org/abs/hep-th/0007079} {arXiv:hep-th/0007079} \BibitemShut
  {NoStop}%
\bibitem [{\citenamefont {Guerrero}\ \emph {et~al.}(2010)\citenamefont
  {Guerrero}, \citenamefont {Melfo}, \citenamefont {Pantoja},\ and\
  \citenamefont {Rodriguez}}]{Guerrero2009}%
  \BibitemOpen
  \bibfield  {author} {\bibinfo {author} {\bibfnamefont {R.}~\bibnamefont
  {Guerrero}}, \bibinfo {author} {\bibfnamefont {A.}~\bibnamefont {Melfo}},
  \bibinfo {author} {\bibfnamefont {N.}~\bibnamefont {Pantoja}}, \ and\
  \bibinfo {author} {\bibfnamefont {R.~O.}\ \bibnamefont {Rodriguez}},\ }\href
  {\doibase 10.1103/PhysRevD.81.086004} {\bibfield  {journal} {\bibinfo
  {journal} {Phys. Rev. D}\ }\textbf {\bibinfo {volume} {81}},\ \bibinfo
  {pages} {086004} (\bibinfo {year} {2010})},\ \Eprint
  {http://arxiv.org/abs/0912.0463} {arXiv:0912.0463 [hep-th]} \BibitemShut
  {NoStop}%
\bibitem [{\citenamefont {Ghoroku}\ and\ \citenamefont
  {Nakamura}(2002)}]{Ghoroku2002}%
  \BibitemOpen
  \bibfield  {author} {\bibinfo {author} {\bibfnamefont {K.}~\bibnamefont
  {Ghoroku}}\ and\ \bibinfo {author} {\bibfnamefont {A.}~\bibnamefont
  {Nakamura}},\ }\href {\doibase 10.1103/PhysRevD.65.084017} {\bibfield
  {journal} {\bibinfo  {journal} {Phys. Rev. D}\ }\textbf {\bibinfo {volume}
  {65}},\ \bibinfo {pages} {084017} (\bibinfo {year} {2002})},\ \Eprint
  {http://arxiv.org/abs/hep-th/0106145} {arXiv:hep-th/0106145} \BibitemShut
  {NoStop}%
\bibitem [{\citenamefont {Carena}\ \emph {et~al.}(2003)\citenamefont {Carena},
  \citenamefont {Ponton}, \citenamefont {Tait},\ and\ \citenamefont
  {Wagner}}]{CarenaPontonTaitWagner2003}%
  \BibitemOpen
  \bibfield  {author} {\bibinfo {author} {\bibfnamefont {M.~S.}\ \bibnamefont
  {Carena}}, \bibinfo {author} {\bibfnamefont {E.}~\bibnamefont {Ponton}},
  \bibinfo {author} {\bibfnamefont {T.~M.}\ \bibnamefont {Tait}}, \ and\
  \bibinfo {author} {\bibfnamefont {C.}~\bibnamefont {Wagner}},\ }\href
  {\doibase 10.1103/PhysRevD.67.096006} {\bibfield  {journal} {\bibinfo
  {journal} {Phys.Rev.}\ }\textbf {\bibinfo {volume} {D67}},\ \bibinfo {pages}
  {096006} (\bibinfo {year} {2003})},\ \Eprint
  {http://arxiv.org/abs/hep-ph/0212307} {arXiv:hep-ph/0212307 [hep-ph]}
  \BibitemShut {NoStop}%
\bibitem [{\citenamefont {Davoudiasl}\ \emph {et~al.}(2003)\citenamefont
  {Davoudiasl}, \citenamefont {Hewett},\ and\ \citenamefont
  {Rizzo}}]{DavoudiaslHewettRizzo2003a}%
  \BibitemOpen
  \bibfield  {author} {\bibinfo {author} {\bibfnamefont {H.}~\bibnamefont
  {Davoudiasl}}, \bibinfo {author} {\bibfnamefont {J.}~\bibnamefont {Hewett}},
  \ and\ \bibinfo {author} {\bibfnamefont {T.}~\bibnamefont {Rizzo}},\ }\href
  {\doibase 10.1103/PhysRevD.68.045002} {\bibfield  {journal} {\bibinfo
  {journal} {Phys. Rev. D}\ }\textbf {\bibinfo {volume} {68}},\ \bibinfo
  {pages} {045002} (\bibinfo {year} {2003})},\ \Eprint
  {http://arxiv.org/abs/hep-ph/0212279} {arXiv:hep-ph/0212279 [hep-ph]}
  \BibitemShut {NoStop}%
\bibitem [{\citenamefont {Giovannini}(2002)}]{Giovannini2002}%
  \BibitemOpen
  \bibfield  {author} {\bibinfo {author} {\bibfnamefont {M.}~\bibnamefont
  {Giovannini}},\ }\href {\doibase 10.1103/PhysRevD.65.124019} {\bibfield
  {journal} {\bibinfo  {journal} {Phys. Rev. D}\ }\textbf {\bibinfo {volume}
  {65}},\ \bibinfo {pages} {124019} (\bibinfo {year} {2002})},\ \Eprint
  {http://arxiv.org/abs/hep-th/0204235} {arXiv:hep-th/0204235 [hep-th]}
  \BibitemShut {NoStop}%
\bibitem [{\citenamefont {Liu}\ \emph {et~al.}(2009{\natexlab{a}})\citenamefont
  {Liu}, \citenamefont {Zhao}, \citenamefont {Wei},\ and\ \citenamefont
  {Duan}}]{Liu20090902}%
  \BibitemOpen
  \bibfield  {author} {\bibinfo {author} {\bibfnamefont {Y.-X.}\ \bibnamefont
  {Liu}}, \bibinfo {author} {\bibfnamefont {Z.-H.}\ \bibnamefont {Zhao}},
  \bibinfo {author} {\bibfnamefont {S.-W.}\ \bibnamefont {Wei}}, \ and\
  \bibinfo {author} {\bibfnamefont {Y.-S.}\ \bibnamefont {Duan}},\ }\href
  {\doibase 10.1088/1475-7516/2009/02/003} {\bibfield  {journal} {\bibinfo
  {journal} {J. Cosmol. Astropart. Phys.}\ }\textbf {\bibinfo {volume}
  {0902}},\ \bibinfo {pages} {003} (\bibinfo {year} {2009}{\natexlab{a}})},\
  \Eprint {http://arxiv.org/abs/0901.0782} {arXiv:0901.0782 [hep-th]}
  \BibitemShut {NoStop}%
\bibitem [{\citenamefont {Guo}\ \emph {et~al.}(2013)\citenamefont {Guo},
  \citenamefont {Herrera-Aguilar}, \citenamefont {Liu}, \citenamefont
  {Malagon-Morejon},\ and\ \citenamefont
  {Mora-Luna}}]{GuoHerrera-AguilarLiuMalagon-MorejonMora-Luna2013}%
  \BibitemOpen
  \bibfield  {author} {\bibinfo {author} {\bibfnamefont {H.}~\bibnamefont
  {Guo}}, \bibinfo {author} {\bibfnamefont {A.}~\bibnamefont
  {Herrera-Aguilar}}, \bibinfo {author} {\bibfnamefont {Y.-X.}\ \bibnamefont
  {Liu}}, \bibinfo {author} {\bibfnamefont {D.}~\bibnamefont
  {Malagon-Morejon}}, \ and\ \bibinfo {author} {\bibfnamefont {R.~R.}\
  \bibnamefont {Mora-Luna}},\ }\href {\doibase 10.1103/PhysRevD.87.095011}
  {\bibfield  {journal} {\bibinfo  {journal} {Phys. Rev. D}\ }\textbf {\bibinfo
  {volume} {87}},\ \bibinfo {pages} {095011} (\bibinfo {year} {2013})},\
  \Eprint {http://arxiv.org/abs/1103.2430} {arXiv:1103.2430 [hep-th]}
  \BibitemShut {NoStop}%
\bibitem [{\citenamefont {Herrera-Aguilar}\ \emph {et~al.}(2014)\citenamefont
  {Herrera-Aguilar}, \citenamefont {Rojas},\ and\ \citenamefont
  {Santos-Rodriguez}}]{Herrera-AguilarRojasSantos-Rodriguez2014}%
  \BibitemOpen
  \bibfield  {author} {\bibinfo {author} {\bibfnamefont {A.}~\bibnamefont
  {Herrera-Aguilar}}, \bibinfo {author} {\bibfnamefont {A.~D.}\ \bibnamefont
  {Rojas}}, \ and\ \bibinfo {author} {\bibfnamefont {E.}~\bibnamefont
  {Santos-Rodriguez}},\ }\href {\doibase 10.1140/epjc/s10052-014-2770-1}
  {\bibfield  {journal} {\bibinfo  {journal} {The European Physical Journal C}\
  }\textbf {\bibinfo {volume} {74}},\ \bibinfo {pages} {2770} (\bibinfo {year}
  {2014})},\ \Eprint {http://arxiv.org/abs/1401.0999} {arXiv:1401.0999
  [hep-th]} \BibitemShut {NoStop}%
\bibitem [{\citenamefont {Liu}\ \emph {et~al.}(2008{\natexlab{b}})\citenamefont
  {Liu}, \citenamefont {Zhang}, \citenamefont {Wei},\ and\ \citenamefont
  {Duan}}]{Liu200808}%
  \BibitemOpen
  \bibfield  {author} {\bibinfo {author} {\bibfnamefont {Y.-X.}\ \bibnamefont
  {Liu}}, \bibinfo {author} {\bibfnamefont {L.-D.}\ \bibnamefont {Zhang}},
  \bibinfo {author} {\bibfnamefont {S.-W.}\ \bibnamefont {Wei}}, \ and\
  \bibinfo {author} {\bibfnamefont {Y.-S.}\ \bibnamefont {Duan}},\ }\href
  {\doibase 10.1088/1126-6708/2008/08/041} {\bibfield  {journal} {\bibinfo
  {journal} {JHEP}\ }\textbf {\bibinfo {volume} {08}},\ \bibinfo {pages} {041}
  (\bibinfo {year} {2008}{\natexlab{b}})},\ \Eprint
  {http://arxiv.org/abs/0803.0098} {arXiv:0803.0098 [hep-th]} \BibitemShut
  {NoStop}%
\bibitem [{\citenamefont {Liu}\ \emph {et~al.}(2012)\citenamefont {Liu},
  \citenamefont {Fu}, \citenamefont {Guo},\ and\ \citenamefont
  {Li}}]{LiuFuGuoLi2012}%
  \BibitemOpen
  \bibfield  {author} {\bibinfo {author} {\bibfnamefont {Y.-X.}\ \bibnamefont
  {Liu}}, \bibinfo {author} {\bibfnamefont {C.-E.}\ \bibnamefont {Fu}},
  \bibinfo {author} {\bibfnamefont {H.}~\bibnamefont {Guo}}, \ and\ \bibinfo
  {author} {\bibfnamefont {H.-T.}\ \bibnamefont {Li}},\ }\href {\doibase
  10.1103/PhysRevD.85.084023} {\bibfield  {journal} {\bibinfo  {journal}
  {Phys.Rev.}\ }\textbf {\bibinfo {volume} {D85}},\ \bibinfo {pages} {084023}
  (\bibinfo {year} {2012})},\ \Eprint {http://arxiv.org/abs/1102.4500}
  {arXiv:1102.4500 [hep-th]} \BibitemShut {NoStop}%
\bibitem [{\citenamefont {Kehagias}\ and\ \citenamefont
  {Tamvakis}(2001)}]{Kehagias2001504}%
  \BibitemOpen
  \bibfield  {author} {\bibinfo {author} {\bibfnamefont {A.}~\bibnamefont
  {Kehagias}}\ and\ \bibinfo {author} {\bibfnamefont {K.}~\bibnamefont
  {Tamvakis}},\ }\href {\doibase 10.1016/S0370-2693(01)00274-X} {\bibfield
  {journal} {\bibinfo  {journal} {Phys. Lett. B}\ }\textbf {\bibinfo {volume}
  {504}},\ \bibinfo {pages} {38} (\bibinfo {year} {2001})},\ \Eprint
  {http://arxiv.org/abs/hep-th/0010112} {arXiv:hep-th/0010112} \BibitemShut
  {NoStop}%
\bibitem [{\citenamefont {Cruz}\ \emph {et~al.}(2010)\citenamefont {Cruz},
  \citenamefont {Tahim},\ and\ \citenamefont {Almeida}}]{CruzTahimAlmeida2010}%
  \BibitemOpen
  \bibfield  {author} {\bibinfo {author} {\bibfnamefont {W.}~\bibnamefont
  {Cruz}}, \bibinfo {author} {\bibfnamefont {M.}~\bibnamefont {Tahim}}, \ and\
  \bibinfo {author} {\bibfnamefont {C.}~\bibnamefont {Almeida}},\ }\href
  {\doibase 10.1016/j.physletb.2010.02.064} {\bibfield  {journal} {\bibinfo
  {journal} {Phys. Lett. B}\ }\textbf {\bibinfo {volume} {686}},\ \bibinfo
  {pages} {259} (\bibinfo {year} {2010})}\BibitemShut {NoStop}%
\bibitem [{\citenamefont {Alencar}\ \emph {et~al.}(2010)\citenamefont
  {Alencar}, \citenamefont {Landim}, \citenamefont {Tahim}, \citenamefont
  {Muniz},\ and\ \citenamefont
  {Costa~Filho}}]{AlencarLandimTahimMunizCosta2010}%
  \BibitemOpen
  \bibfield  {author} {\bibinfo {author} {\bibfnamefont {G.}~\bibnamefont
  {Alencar}}, \bibinfo {author} {\bibfnamefont {R.}~\bibnamefont {Landim}},
  \bibinfo {author} {\bibfnamefont {M.}~\bibnamefont {Tahim}}, \bibinfo
  {author} {\bibfnamefont {C.}~\bibnamefont {Muniz}}, \ and\ \bibinfo {author}
  {\bibfnamefont {R.}~\bibnamefont {Costa~Filho}},\ }\href {\doibase
  10.1016/j.physletb.2010.09.005} {\bibfield  {journal} {\bibinfo  {journal}
  {Phys. Lett. B}\ }\textbf {\bibinfo {volume} {693}},\ \bibinfo {pages} {503}
  (\bibinfo {year} {2010})},\ \Eprint {http://arxiv.org/abs/1008.0678}
  {arXiv:1008.0678 [hep-th]} \BibitemShut {NoStop}%
\bibitem [{\citenamefont {Cruz}\ \emph
  {et~al.}(2013{\natexlab{a}})\citenamefont {Cruz}, \citenamefont {Lima},\ and\
  \citenamefont {Almeida}}]{CruzLimaAlmeida2013}%
  \BibitemOpen
  \bibfield  {author} {\bibinfo {author} {\bibfnamefont {W.}~\bibnamefont
  {Cruz}}, \bibinfo {author} {\bibfnamefont {A.~R.}\ \bibnamefont {Lima}}, \
  and\ \bibinfo {author} {\bibfnamefont {C.}~\bibnamefont {Almeida}},\ }\href
  {\doibase 10.1103/PhysRevD.87.045018} {\bibfield  {journal} {\bibinfo
  {journal} {Phys. Rev. D}\ }\textbf {\bibinfo {volume} {87}},\ \bibinfo
  {pages} {045018} (\bibinfo {year} {2013}{\natexlab{a}})},\ \Eprint
  {http://arxiv.org/abs/1211.7355} {arXiv:1211.7355 [hep-th]} \BibitemShut
  {NoStop}%
\bibitem [{\citenamefont {Fu}\ \emph {et~al.}(2011)\citenamefont {Fu},
  \citenamefont {Liu},\ and\ \citenamefont {Guo}}]{FuLiuGuo2011}%
  \BibitemOpen
  \bibfield  {author} {\bibinfo {author} {\bibfnamefont {C.-E.}\ \bibnamefont
  {Fu}}, \bibinfo {author} {\bibfnamefont {Y.-X.}\ \bibnamefont {Liu}}, \ and\
  \bibinfo {author} {\bibfnamefont {H.}~\bibnamefont {Guo}},\ }\href {\doibase
  10.1103/PhysRevD.84.044036} {\bibfield  {journal} {\bibinfo  {journal} {Phys.
  Rev. D}\ }\textbf {\bibinfo {volume} {84}},\ \bibinfo {pages} {044036}
  (\bibinfo {year} {2011})},\ \Eprint {http://arxiv.org/abs/1101.0336}
  {arXiv:1101.0336 [hep-th]} \BibitemShut {NoStop}%
\bibitem [{\citenamefont {Cruz}\ \emph {et~al.}(2009)\citenamefont {Cruz},
  \citenamefont {Tahim},\ and\ \citenamefont {Almeida}}]{CruzTahimAlmeida2009}%
  \BibitemOpen
  \bibfield  {author} {\bibinfo {author} {\bibfnamefont {W.}~\bibnamefont
  {Cruz}}, \bibinfo {author} {\bibfnamefont {M.}~\bibnamefont {Tahim}}, \ and\
  \bibinfo {author} {\bibfnamefont {C.}~\bibnamefont {Almeida}},\ }\href
  {\doibase 10.1209/0295-5075/88/41001} {\bibfield  {journal} {\bibinfo
  {journal} {Europhys.Lett.}\ }\textbf {\bibinfo {volume} {88}},\ \bibinfo
  {pages} {41001} (\bibinfo {year} {2009})},\ \Eprint
  {http://arxiv.org/abs/0912.1029} {arXiv:0912.1029 [hep-th]} \BibitemShut
  {NoStop}%
\bibitem [{\citenamefont {Christiansen}\ \emph {et~al.}(2010)\citenamefont
  {Christiansen}, \citenamefont {Cunha},\ and\ \citenamefont
  {Tahim}}]{ChristiansenCunhaTahim2010}%
  \BibitemOpen
  \bibfield  {author} {\bibinfo {author} {\bibfnamefont {H.}~\bibnamefont
  {Christiansen}}, \bibinfo {author} {\bibfnamefont {M.}~\bibnamefont {Cunha}},
  \ and\ \bibinfo {author} {\bibfnamefont {M.}~\bibnamefont {Tahim}},\ }\href
  {\doibase 10.1103/PhysRevD.82.085023} {\bibfield  {journal} {\bibinfo
  {journal} {Phys. Rev. D}\ }\textbf {\bibinfo {volume} {82}},\ \bibinfo
  {pages} {085023} (\bibinfo {year} {2010})},\ \Eprint
  {http://arxiv.org/abs/1006.1366} {arXiv:1006.1366 [hep-th]} \BibitemShut
  {NoStop}%
\bibitem [{\citenamefont {Cruz}\ \emph
  {et~al.}(2013{\natexlab{b}})\citenamefont {Cruz}, \citenamefont {Maluf},\
  and\ \citenamefont {Almeida}}]{CruzMalufAlmeida2013}%
  \BibitemOpen
  \bibfield  {author} {\bibinfo {author} {\bibfnamefont {W.}~\bibnamefont
  {Cruz}}, \bibinfo {author} {\bibfnamefont {R.}~\bibnamefont {Maluf}}, \ and\
  \bibinfo {author} {\bibfnamefont {C.}~\bibnamefont {Almeida}},\ }\href@noop
  {} {\  (\bibinfo {year} {2013}{\natexlab{b}})},\ \Eprint
  {http://arxiv.org/abs/1303.1096} {arXiv:1303.1096 [hep-th]} \BibitemShut
  {NoStop}%
\bibitem [{\citenamefont {Chumbes}\ \emph {et~al.}(2012)\citenamefont
  {Chumbes}, \citenamefont {Hoff~da Silva},\ and\ \citenamefont
  {Hott}}]{ChumbesHoffHott2012}%
  \BibitemOpen
  \bibfield  {author} {\bibinfo {author} {\bibfnamefont {A.}~\bibnamefont
  {Chumbes}}, \bibinfo {author} {\bibfnamefont {J.}~\bibnamefont {Hoff~da
  Silva}}, \ and\ \bibinfo {author} {\bibfnamefont {M.}~\bibnamefont {Hott}},\
  }\href {\doibase 10.1103/PhysRevD.85.085003} {\bibfield  {journal} {\bibinfo
  {journal} {Phys. Rev. D}\ }\textbf {\bibinfo {volume} {85}},\ \bibinfo
  {pages} {085003} (\bibinfo {year} {2012})},\ \Eprint
  {http://arxiv.org/abs/1108.3821} {arXiv:1108.3821 [hep-th]} \BibitemShut
  {NoStop}%
\bibitem [{\citenamefont {Bazeia}\ and\ \citenamefont
  {Gomes}(2004)}]{Bazeia200405}%
  \BibitemOpen
  \bibfield  {author} {\bibinfo {author} {\bibfnamefont {D.}~\bibnamefont
  {Bazeia}}\ and\ \bibinfo {author} {\bibfnamefont {A.~R.}\ \bibnamefont
  {Gomes}},\ }\href {\doibase 10.1088/1126-6708/2004/05/012} {\bibfield
  {journal} {\bibinfo  {journal} {JHEP}\ }\textbf {\bibinfo {volume} {05}},\
  \bibinfo {pages} {012} (\bibinfo {year} {2004})},\ \Eprint
  {http://arxiv.org/abs/hep-th/0403141} {arXiv:hep-th/0403141} \BibitemShut
  {NoStop}%
\bibitem [{\citenamefont {de~Souza~Dutra}\ \emph {et~al.}(2008)\citenamefont
  {de~Souza~Dutra}, \citenamefont {A.~C. Amaro~de Faria},\ and\ \citenamefont
  {Hott}}]{SouzaDutra200878}%
  \BibitemOpen
  \bibfield  {author} {\bibinfo {author} {\bibfnamefont {A.}~\bibnamefont
  {de~Souza~Dutra}}, \bibinfo {author} {\bibfnamefont {J.}~\bibnamefont {A.~C.
  Amaro~de Faria}}, \ and\ \bibinfo {author} {\bibfnamefont {M.}~\bibnamefont
  {Hott}},\ }\href {\doibase 10.1103/PhysRevD.78.043526} {\bibfield  {journal}
  {\bibinfo  {journal} {Phys. Rev. D}\ }\textbf {\bibinfo {volume} {78}},\
  \bibinfo {pages} {043526} (\bibinfo {year} {2008})},\ \Eprint
  {http://arxiv.org/abs/0807.0586} {arXiv:0807.0586 [hep-th]} \BibitemShut
  {NoStop}%
\bibitem [{\citenamefont {Gomes}(2006)}]{Gomes2006}%
  \BibitemOpen
  \bibfield  {author} {\bibinfo {author} {\bibfnamefont {A.~R.}\ \bibnamefont
  {Gomes}},\ }\href@noop {} {\enquote {\bibinfo {title} {Gravity on the bloch
  brane},}\ } (\bibinfo {year} {2006}),\ \Eprint
  {http://arxiv.org/abs/hep-th/0611291} {arXiv:hep-th/0611291} \BibitemShut
  {NoStop}%
\bibitem [{\citenamefont {Correa}\ \emph {et~al.}(2011)\citenamefont {Correa},
  \citenamefont {de~Souza~Dutra},\ and\ \citenamefont
  {Hott}}]{CorreaDutraHott2010}%
  \BibitemOpen
  \bibfield  {author} {\bibinfo {author} {\bibfnamefont {R.}~\bibnamefont
  {Correa}}, \bibinfo {author} {\bibfnamefont {A.}~\bibnamefont
  {de~Souza~Dutra}}, \ and\ \bibinfo {author} {\bibfnamefont {M.}~\bibnamefont
  {Hott}},\ }\href {\doibase 10.1088/0264-9381/28/15/155012} {\bibfield
  {journal} {\bibinfo  {journal} {Class.Quant.Grav.}\ }\textbf {\bibinfo
  {volume} {28}},\ \bibinfo {pages} {155012} (\bibinfo {year} {2011})},\
  \Eprint {http://arxiv.org/abs/1011.1849} {arXiv:1011.1849 [hep-th]}
  \BibitemShut {NoStop}%
\bibitem [{\citenamefont {Xie}\ \emph {et~al.}(2013)\citenamefont {Xie},
  \citenamefont {Yang},\ and\ \citenamefont {Zhao}}]{XieYangZhao2013}%
  \BibitemOpen
  \bibfield  {author} {\bibinfo {author} {\bibfnamefont {Q.-Y.}\ \bibnamefont
  {Xie}}, \bibinfo {author} {\bibfnamefont {J.}~\bibnamefont {Yang}}, \ and\
  \bibinfo {author} {\bibfnamefont {L.}~\bibnamefont {Zhao}},\ }\href {\doibase
  10.1103/PhysRevD.88.105014} {\bibfield  {journal} {\bibinfo  {journal}
  {Phys.Rev.}\ }\textbf {\bibinfo {volume} {D88}},\ \bibinfo {pages} {105014}
  (\bibinfo {year} {2013})},\ \Eprint {http://arxiv.org/abs/1310.4585}
  {arXiv:1310.4585 [hep-th]} \BibitemShut {NoStop}%
\bibitem [{Note1()}]{Note1}%
  \BibitemOpen
  \bibinfo {note} {The details of this function can be found in the website:
  http://mathworld.wolfram.com/CompleteEllipticIntegraloftheFirstKind.html}\BibitemShut
  {NoStop}%
\bibitem [{\citenamefont {Bazeia}\ \emph {et~al.}(2004)\citenamefont {Bazeia},
  \citenamefont {Furtado},\ and\ \citenamefont {Gomes}}]{Bazeia2004}%
  \BibitemOpen
  \bibfield  {author} {\bibinfo {author} {\bibfnamefont {D.}~\bibnamefont
  {Bazeia}}, \bibinfo {author} {\bibfnamefont {C.}~\bibnamefont {Furtado}}, \
  and\ \bibinfo {author} {\bibfnamefont {A.~R.}\ \bibnamefont {Gomes}},\ }\href
  {\doibase 10.1088/1475-7516/2004/02/002} {\bibfield  {journal} {\bibinfo
  {journal} {J. Cosmol. Astropart. Phys.}\ }\textbf {\bibinfo {volume}
  {0402}},\ \bibinfo {pages} {002} (\bibinfo {year} {2004})},\ \Eprint
  {http://arxiv.org/abs/hep-th/0308034} {arXiv:hep-th/0308034} \BibitemShut
  {NoStop}%
\bibitem [{\citenamefont {Liu}\ \emph {et~al.}(2009{\natexlab{b}})\citenamefont
  {Liu}, \citenamefont {Yang}, \citenamefont {Zhao}, \citenamefont {Fu},\ and\
  \citenamefont {Duan}}]{Liu2009}%
  \BibitemOpen
  \bibfield  {author} {\bibinfo {author} {\bibfnamefont {Y.-X.}\ \bibnamefont
  {Liu}}, \bibinfo {author} {\bibfnamefont {J.}~\bibnamefont {Yang}}, \bibinfo
  {author} {\bibfnamefont {Z.-H.}\ \bibnamefont {Zhao}}, \bibinfo {author}
  {\bibfnamefont {C.-E.}\ \bibnamefont {Fu}}, \ and\ \bibinfo {author}
  {\bibfnamefont {Y.-S.}\ \bibnamefont {Duan}},\ }\href {\doibase
  10.1103/PhysRevD.80.065019} {\bibfield  {journal} {\bibinfo  {journal} {Phys.
  Rev. D}\ }\textbf {\bibinfo {volume} {80}},\ \bibinfo {pages} {065019}
  (\bibinfo {year} {2009}{\natexlab{b}})},\ \Eprint
  {http://arxiv.org/abs/0904.1785} {arXiv:0904.1785 [hep-th]} \BibitemShut
  {NoStop}%
\bibitem [{\citenamefont {Weinberg}(1995)}]{Weinberg1995}%
  \BibitemOpen
  \bibfield  {author} {\bibinfo {author} {\bibfnamefont {S.}~\bibnamefont
  {Weinberg}},\ }\href@noop {} {\emph {\bibinfo {title} {Quantum theory of
  fields}}},\ Vol.\ \bibinfo {volume} {1. Foundations}\ (\bibinfo  {publisher}
  {Press Syndicate of the University of Cambridge},\ \bibinfo {year}
  {1995})\BibitemShut {NoStop}%
\end{thebibliography}%

\end{document}